\documentclass[11pt]{article}

\usepackage{graphicx}
\usepackage{xspace}

\usepackage{amsthm}
\usepackage{amssymb}
\usepackage{amsmath}
\usepackage{algorithm}
\usepackage{algpseudocode}
\usepackage{caption}
\DeclareCaptionType{copyrightbox}
\usepackage{subcaption}
\usepackage{paralist}
\usepackage{booktabs}
\usepackage{pifont}
\usepackage{times}
\usepackage{dsfont}
\usepackage{url}
\usepackage{multirow,bigdelim}
\usepackage{hyphenat}
\usepackage[T1]{fontenc}
\usepackage{rotating}
\usepackage{pifont}
\usepackage{arydshln}
\usepackage[margin=1in]{geometry}

\newtheorem{insight}{Insight}

\newtheorem{lemma}{Lemma}

\newtheorem*{keyinsight}{Key Insight}

\newcommand{\hide}[1]{}

\newcommand{\bit}{\begin{compactitem}}
\newcommand{\eit}{\end{compactitem}}
\newcommand{\ben}{\begin{compactenum}}
\newcommand{\een}{\end{compactenum}}

\newcommand\textvtt[1]{{\normalfont\fontfamily{cmvtt}\selectfont #1}}

\newcommand{\fastfollowerz}{\textvtt{fastfollowerz}\xspace}
\newcommand{\plusfollower}{\textvtt{plusfollower}\xspace}
\newcommand{\devumi}{\textvtt{devumi}\xspace}
\newcommand{\hitfollow}{\textvtt{hitfollow}\xspace}
\newcommand{\newfollow}{\textvtt{newfollow}\xspace}
\newcommand{\intertwitter}{\textvtt{intertwitter}\xspace}
\newcommand{\twitterboost}{\textvtt{twitterboost}\xspace}
\newcommand{\bigfolo}{\textvtt{bigfolo}\xspace}

\begin{document}

\title{The Many Faces of Link Fraud}
\date{}

\author{Neil Shah$^*$, Hemank Lamba$^*$, Alex Beutel$^\dagger$, Christos Faloutsos$^*$ \\
Carnegie Mellon University$^*$, Google Research$^\dagger$\\
Email: \{nshah, hlamba, christos\}@cs.cmu.edu, alexbeutel@google.com
}

\maketitle

\begin{abstract}

Most past work on social network link fraud detection tries to separate genuine users from fraudsters, implicitly assuming that there is only one type of fraudulent behavior. But is this assumption true?  And, in either case, what are the characteristics of such fraudulent behaviors?  In this work, we set up \emph{honeypots}, (``dummy'' social network accounts), and buy fake followers (after careful IRB approval).  We report the signs of such behaviors including oddities in local network connectivity, account attributes, and similarities and differences across fraud providers. 
Most valuably, we discover and characterize {\em several} types of fraud behaviors. We discuss how to leverage our insights in practice by engineering strongly performing entropy-based features and demonstrating high classification accuracy.  Our contributions are (a) \emph{instrumentation}: we detail our experimental setup and carefully engineered data collection process to scrape Twitter data while respecting API rate-limits, (b) \emph{observations on fraud multimodality}: we analyze our honeypot fraudster ecosystem and give surprising insights into the multifaceted behaviors of these fraudster types, and (c) \emph{features}: we propose novel features that give strong (\emph{>0.95} precision/recall) discriminative power on ground-truth Twitter data.

\end{abstract}
\section{Introduction}

What are the characteristics of fraudulent accounts in online social networks?  Understanding the behavior and actions of fraudsters is paramount to building effective anti-fraud algorithms.  While previous works in social network fraud detection have primarily focused on leveraging signature properties of fraudsters including temporally synchronized behavior \cite{beutel2013copycatch}, excessively dense \cite{prakash2010eigenspokes} and oddly distributed \cite{shah2014spotting} graph connectivity, uncommon account names \cite{freeman2013using} and spammy links \cite{grier2010spam}, our work focuses on establishing the veracity and applicability of these assumptions.  In doing so, we ask: do all fraudsters share the same signature behavior, or are there multiple signatures?  Since fraud detection is an adversarial setting in which fraudsters are constantly adapting to in-place detection mechanisms, it is important to constantly monitor and evaluate the strategies that fraudsters are employing to profitably perform ingenuine actions to better inform future detection mechanisms.  

We focus on one particular setting of social network fraud called \emph{link fraud} which involves the use of fake, \emph{sockpuppet} accounts to create links, or graph connections, which represent followership or support of target, customer entities.  Fake links artificially inflate the follower count of customer accounts, making them appear more popular than they actually are.  These fake links are deceptive to authentic users and hinder the performance of machine learning algorithms which rely on authentic user input to recommend relevant and useful content to their userbase.

To study the behavior of these fake follower accounts, we employ the use of \emph{honeypots}, or dummy accounts on which we solicit fake Twitter followers sourced from various fraud service providers. Honeypots enable us to have an clear signal of fake follower activity which is not tainted by follows from real accounts.  Upon setting up the honeypot accounts and purchasing fake followers, we instrument a number of carefully engineered tracking scripts which poll Twitter API to store details including account relationships and attributes over a period of time.  This allows us to collect a rich representation of the fraudster ecosystem which we subsequently analyze.

In this work, we make and explore the following key observation:

\begin{keyinsight}[Fraud Multimodality]
 There are multiple types of link fraud which exhibit notably different network structures and patterns in account attribute settings.
\end{keyinsight}

\begin{figure*}[t!]
  \centering
  \begin{tabular}{c|cc|c}
   \textbf{Gen. users} & \textbf{Pre. fraud} & \textbf{Fre. fraud} & \\
   \begin{subfigure}[t]{0.16\textwidth}
    \centering
    \includegraphics[width=\textwidth]{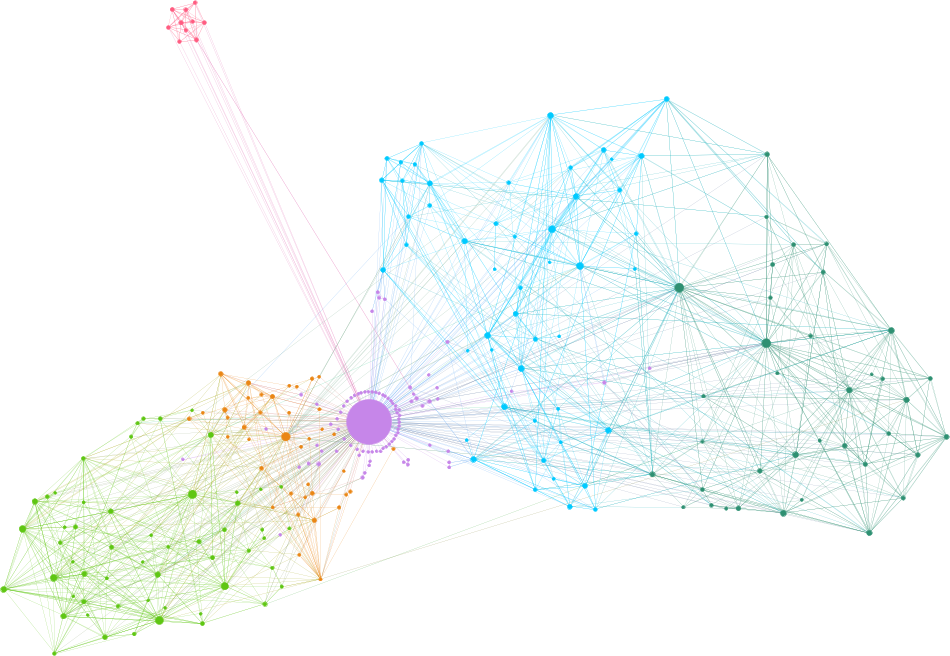}
    \caption[Genuine user egonet visualization]{Visualization}
    \label{fig:crown_honest}
   \end{subfigure}
  &
   \begin{subfigure}[t]{0.16\textwidth}
     \centering
     \includegraphics[width=\textwidth]{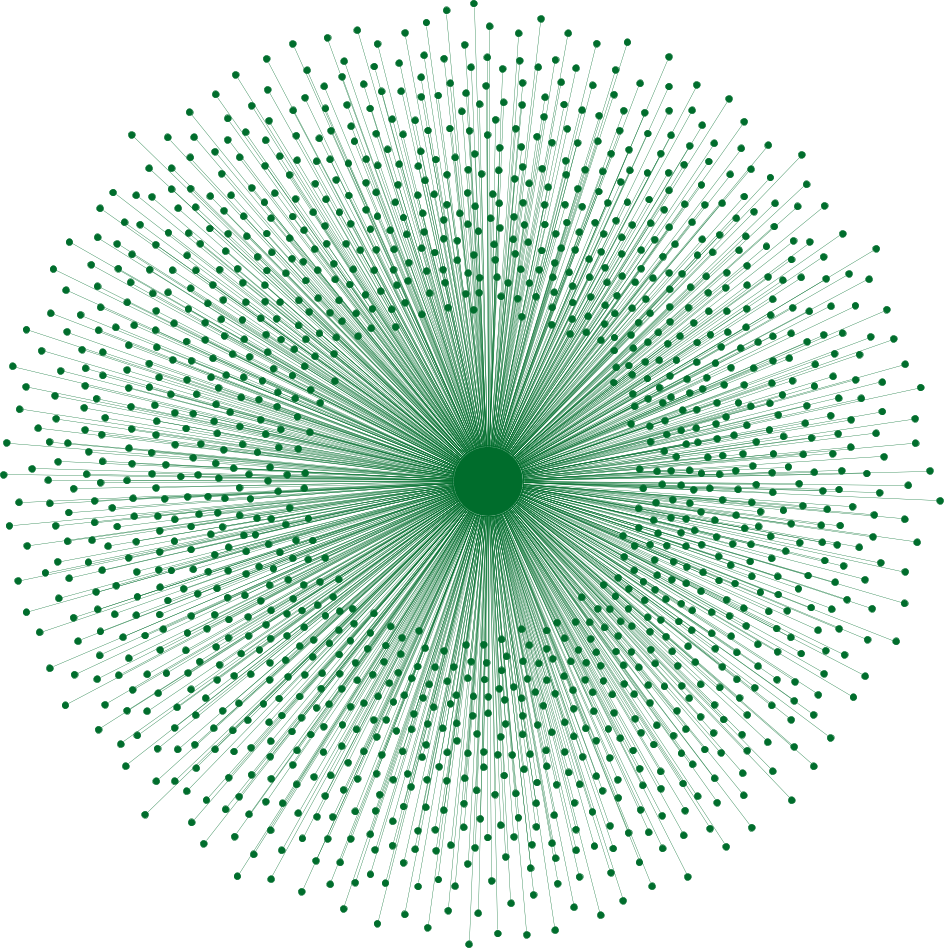}
     \caption[Premium fraudster egonet visualization]{Visualization}
     \label{fig:crown_premium}
    \end{subfigure}
  &
   \begin{subfigure}[t]{0.16\textwidth}
    \centering
    \includegraphics[width=\textwidth]{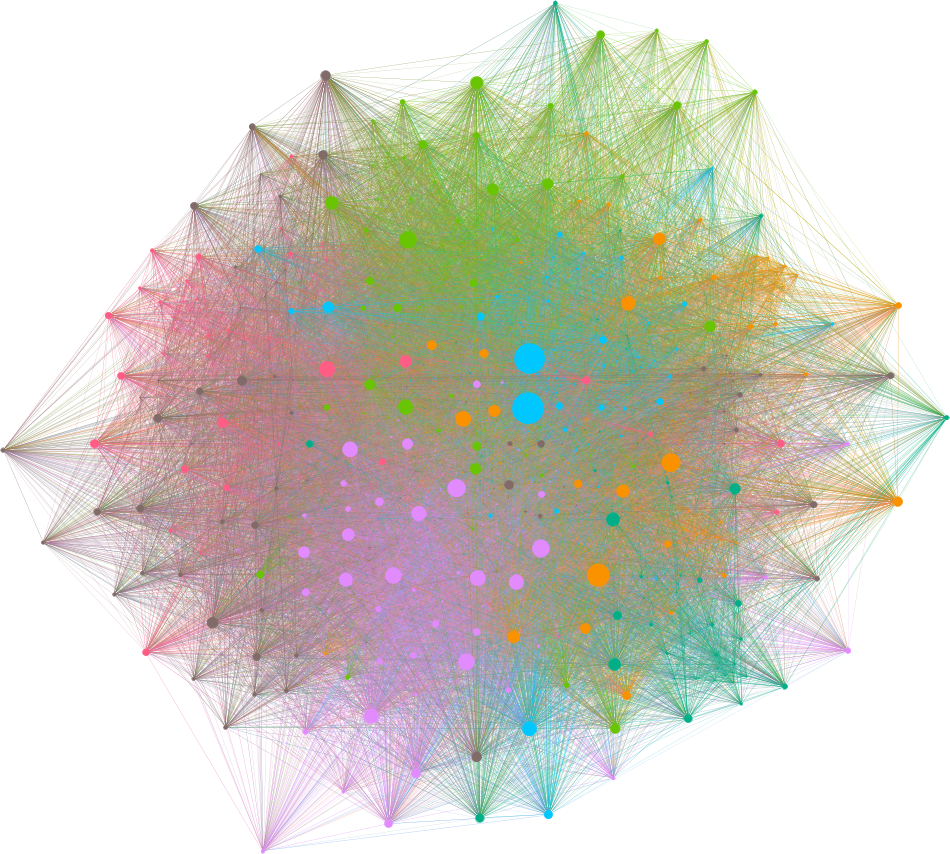}
    \caption[Freemium fraudster egonet visualization]{Visualization}
    \label{fig:crown_freemium}
   \end{subfigure}
  & 
  \multirow{2}{*}[0.85in]{ 
  \begin{subfigure}{0.49\textwidth}
    \centering
    \includegraphics[width=\textwidth]{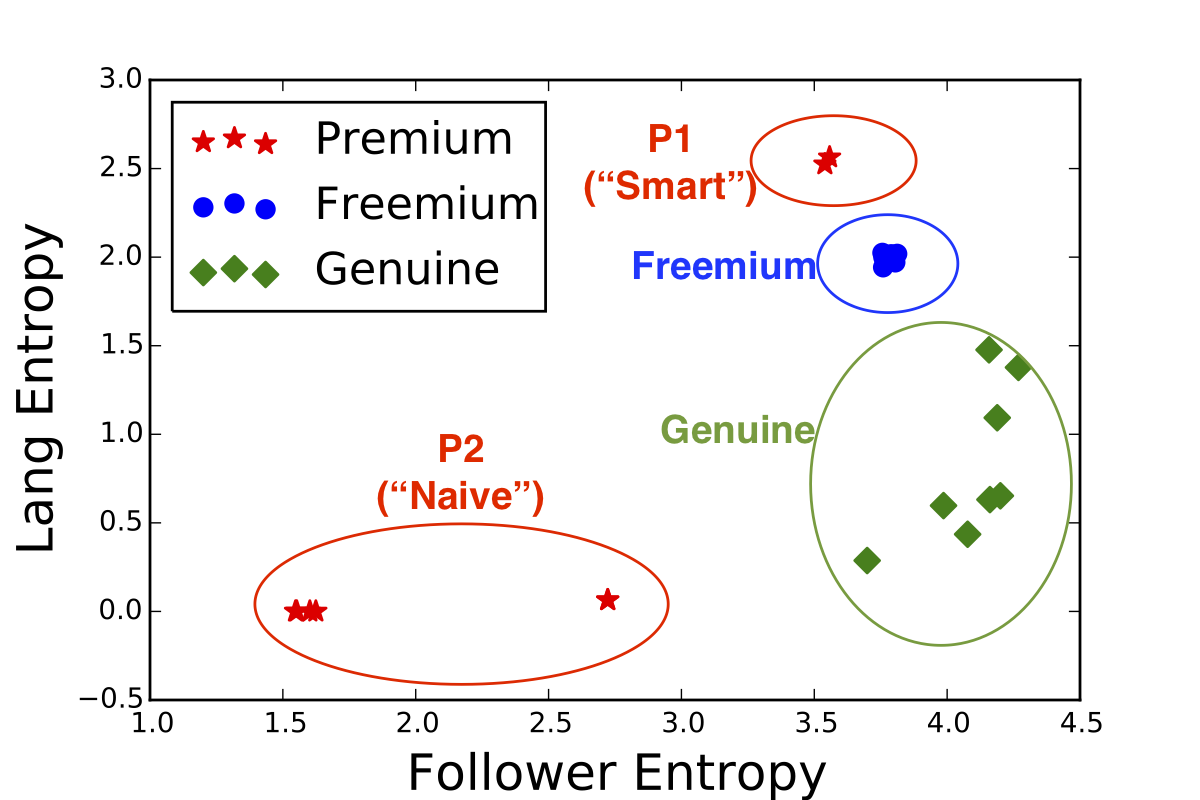}
    \caption[Diverse attribute behaviors across user types]{Diverse attribute behavior}
    \label{fig:crown_attributes}
  \end{subfigure}
  } \\ \bigskip
   \begin{subfigure}[t]{0.16\textwidth}
    \centering
    \includegraphics[width=\textwidth]{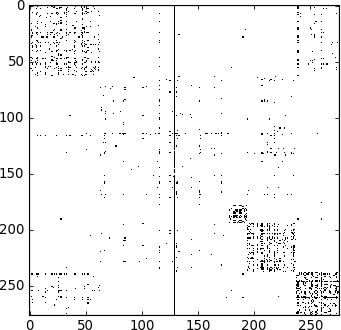}
    \caption[Genuine user egonet adjacency matrix]{Adjacency}
    \label{fig:crown_honest_spy}
   \end{subfigure}
  &
   \begin{subfigure}[t]{0.16\textwidth}
     \centering
     \includegraphics[width=\textwidth]{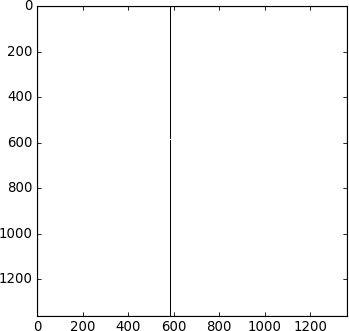}
     \caption[Premium fraudster egonet adjacency matrix]{Adjacency}
     \label{fig:crown_premium_spy}
    \end{subfigure}
  &
   \begin{subfigure}[t]{0.16\textwidth}
    \centering
    \includegraphics[width=\textwidth]{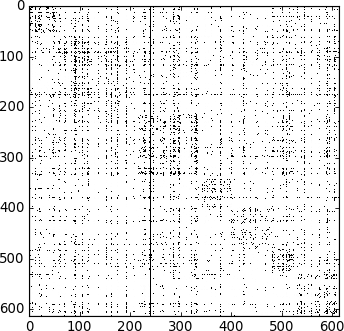}
    \caption[Freemium fraudster egonet adjacency matrix]{Adjacency}
    \label{fig:crown_freemium_spy}
   \end{subfigure}
    &
  \end{tabular}
  \caption[Comparison of network and attribute features across user types]{\textbf{Freemium (Fre) and premium (Pre) fraud types have different local network structure and account attributes compared to genuine behavior.}  Nodes are colored by modularity class, and sized proportional to in-degree in (a)-(c).  The associated, reordered adjacency matrices are shown in (e)-(g) -- the vertical line in each spyplot indicates the the central node.  Notice the block community structure in genuine followers compared to the star structure for premium and near-clique structure for freemium followers.  (d) shows differences in attribute (language and follower) entropy over the various behaviors, showing how fraud patterns skew attribute distributions away from genuine ones.}
  \label{fig:crown_oec}
\end{figure*}

Specifically, we focus on studying and characterizing the network connectivity properties and attribute distributions which are exhibited by fake followers involved in these different types of fraud.  We detail a number of further observations on how these types of behavior induce different, odd network structures and suspicious patterns in account attributes.  Figure \ref{fig:crown_oec} shows the contrast in follower connectivity of a genuine account versus two distinct types of fraudsters.  Through our analysis, we additionally engineer strong features which enable us to discriminate these fraudulent users from genuine ones using novel (first-order) follower entropy features.  

Summarily, our work offers the following notable contributions:

\begin{compactitem}
 \item \textbf{Instrumentation}: We detail our experimental setup and data scraping tools which gather a wealth of Twitter user information while respecting API rate limits.
 \item \textbf{Observations on Fraud Multimodality}: We discover that link fraud is not unimodal and instead has multiple types, and identify and characterize two such types: \emph{freemium} and \emph{premium}, with the possibility of more.
 \item \textbf{Features}: Based on the above observations, we carefully engineer novel, entropy-based features which allow us to accurately discern fraudsters from genuine users in our ground-truth Twitter dataset with near-perfect F1-score.
\end{compactitem}

\section{Related Work}

We categorize related work into two categories: underground market studies and fraud detection approaches.

\vspace{1mm}
\noindent \textbf{Underground Markets:} Prior works have shown the use of fake accounts for followers in social media \cite{Thomas:2013}, phone-verified email accounts \cite{Thomas:2014}, Facebook likes \cite{beutel2013copycatch}, etc. These accounts are often used to spread spam \cite{grier2010spam,Gao:2010} and misinformation \cite{Gupta:13,Gupta:2013}. \cite{TwitterWebImpact:13} estimates that the fake follower market produces \$360 million per year.   Recently, several works have studied the existence of underground online markets where these fraudulent actions can be purchased --  
\cite{Wang:2012, Motoyama:2010} explore underground markets providing fake content, reviews and solutions to security mechanisms. \cite{Thomas:2013} studies several fraud providers over time and describes trends in pricing, account names and IP diversity.  \cite{Stringhini:2013} compares growth rates of accounts with legitimate and fraudulent followers. \cite{Aggarwal:2015} observes the varying retention and reliability of various fraud providers. 
Comparatively, our work is the first to identify major social graph differences between fraud types and across providers, and propose novel entropy-based features for capturing these behaviors.

\vspace{1mm}
\noindent \textbf{Fraud Detection:} \cite{Benevenuto:2010, Lee:2010} use profile features to detect spammers on Twitter. \cite{Stringhini:2010} passively analyzes accounts with promiscuous following behavior and builds a classifier using profile and messaging features. \cite{Cai:2012, yu2006sybilguard} aim to find fake accounts in social networks via a generative stochastic model and a random-walk based method respectively -- both assume small cuts between fake and genuine nodes. \cite{beutel2013copycatch, cao2014uncovering} use graph-traversal based methods to find users with temporally synchronized actions on Facebook. \cite{shah2014spotting, jiang2014catchsync, prakash2010eigenspokes} propose spectral methods which identify dense or odd graph structures indicative of fraud.  

\section{Know Thy Enemy: Characterizing Link Fraud}
In this section, we discuss some preliminaries about instrumentation, data collection and relevant metrics, and next illustrate numerous insights about network connectivity and account attributes of link fraudsters.
\subsection{Setup and Data Collection}

We first discuss how we identified and purchased followers from target fraud service providers, and next detail the scraping task, followed by preliminaries.

\begin{table}[t!]
\renewcommand{\arraystretch}{0.8}
\setlength\tabcolsep{4pt}
\centering
\caption[Honeypot account summaries]{Honeypot account summaries.}
\label{tbl:hp_acc_summaries}
\begin{tabular}{@{}llllll@{}}
\toprule
\textbf{Service}  & \textbf{Type} & \textbf{Cost}  & \textbf{\begin{tabular}[c]{@{}l@{}}Followers\\ bought\end{tabular}} & \textbf{\begin{tabular}[c]{@{}l@{}}Followers \\ delivered\end{tabular}} & \textbf{\begin{tabular}[c]{@{}l@{}}Followers\\ remaining\end{tabular}} \\ \midrule
\multirow{2}{*}{\fastfollowerz} & \multirow{2}{*}{Premium} & \multirow{2}{*}{\$19}               & \multirow{2}{*}{1000}                                                                & 1060                                                                    & 1059                                                                   \\
&	&	& 	& 1060                                                                    & 1059                                                                   \\
\multirow{2}{*}{\intertwitter}  & \multirow{2}{*}{Premium}  & \multirow{2}{*}{\$14}               & \multirow{2}{*}{1000}                                                                & 1099                                                                    & 977                                                                    \\
&	&	&       & 1102                                                                    & 974                                                                    \\
\multirow{2}{*}{\devumi}   & \multirow{2}{*}{Premium}      & \multirow{2}{*}{\$19}               & \multirow{2}{*}{1000}                                                                & 1360                                                                    & 1358                                                                   \\
&	&	& 	   & 1354                                                                    & 1354                                                                   \\
\multirow{2}{*}{\twitterboost} & \multirow{2}{*}{Premium}  & \multirow{2}{*}{\$12}               & \multirow{2}{*}{1000}                                                                & 1361                                                                    & 1361                                                                   \\
&	&	&	   & 1350         & 1350                                                                   \\
\midrule
\multirow{2}{*}{\plusfollower} & \multirow{2}{*}{Freemium} & \multirow{2}{*}{\textsterling 9.99} & \multirow{2}{*}{1000}                                                                & 1094                                                                    & 748                                                                    \\
&	&	&	& 1078                                                                    & 737                                                                    \\
\multirow{2}{*}{\hitfollow}  & \multirow{2}{*}{Freemium}  & \multirow{2}{*}{\textsterling 9.99} & \multirow{2}{*}{1000}                                                                & 926                                                                     & 623                                                                    \\
&	&	&	& 937                                                                     & 638                                                                    \\
\multirow{2}{*}{\newfollow}  & \multirow{2}{*}{Freemium}  & \multirow{2}{*}{\textsterling 9.99} & \multirow{2}{*}{1000}                                                                & 884                                                                     & 600                                                                    \\
&	&	&	& 883                                                                     & 589                                                                    \\
\multirow{2}{*}{\bigfolo}   & \multirow{2}{*}{Freemium}    & \multirow{2}{*}{\textsterling 9.99} & \multirow{2}{*}{1000}                                                                & 872                                                                     & 594                                                                    \\
&	&	&	& 865                                                                     & 577                                                                    \\ \bottomrule
\end{tabular}
\end{table}

\subsubsection{Purchasing Fake Followers}

There are a number of different fraud service providers easily accessible and available on the web. We begin by identifying these services so we can purchase fake followers from them. To identify these services, we used Google search and queried using keywords such as ``buy Twitter followers.''  Combining the search results, we obtained a list of websites which claim to provide these services.

From surveying the websites on this list, we notice there are several prevalent models of service -- we categorize these into two frameworks: \emph{premium} and \emph{freemium}.  Premium services offer customers multiple tiers of follower counts (1K, 5K, 10K, etc.) for various amounts of money and ask only for the customer's Twitter username and a form of payment. Freemium services offer both a paid option as in premium services, but additionally offer a free option which does not ask the user for money, but instead requires the user to provide their Twitter login details to the service.  In return for these details, the services promise to direct a small number of followers to the account. 

We next setup a pool of honeypot accounts by repeating the Twitter account creation process a number of times using monikers from online screenname generators.  We found that to create a sizeable pool of honeypots, we needed to distribute the account creation over several IPs in order to avoid phone verification prompts.  Upon setting up the pool of honeypots, we purchased basic follower packages from several premium and paid freemium services, avoiding rarely used ones with low Alexa rank.  Summarily, we bought 1K followers from 8 different services (4 freemium, 4 premium) to 2 honeypot accounts per service.  We chose to purchase 2 honeypot accounts per service instead of only 1 in order to examine the overlap dynamics of fake links to multiple customers.  The final list of the services we used, service types, costs and their follower counts are summarized Table \ref{tbl:hp_acc_summaries}.  Honeypots were created on the same day, and follower purchases were all done at the same time.  Furthermore, the honeypots attracted no followers by themselves prior to the purchases.  As a result, we posit that all followers of the honeypots are fake.

\subsubsection{Instrumentation Details}

\vspace{1mm}
\noindent \textbf{Reproducibility:} Code available at \texttt{https://goo.gl/qMBWim}.

\vspace{1mm}
We use the REST API to scrape data relevant to our operation from Twitter. As the API heavily rate-limits various data resource types, it is only feasible to extract a limited amount of information as an end-user.  Prior to purchasing fake followers, we start a number of Python scripts which poll the API and insert data into a Postgres database:

\vspace{1mm}

\noindent \textbf{Honeypot account details}: Every hour, we collect public details for each honeypot Twitter account including number of friends and followees, number of favorites, number of Tweets, language, etc. 

\vspace{1mm}
\noindent \textbf{Honeypot account follower IDs}: Every 12 hours, we collect the list of follower IDs for each honeypot.  Since the honeypots were created with empty profiles, we can safely assume that all followers to these accounts were fraudulent and purchased.  

\vspace{1mm}
\noindent \textbf{Honeypot account follower details}: Every day, we extract public details for each of the accounts in the honeypot follower list.  

\vspace{1mm}
\noindent \textbf{Honeypot account followers' friends/followers IDs}: Every day, we collect the list of friend and follower IDs of the honeypot followers to examine their other connectivity. 

\vspace{1mm}
\noindent \textbf{Honeypot account followers' friends/followers details}: Every 3 days, we extract public details for each of the friends and followers of the honeypot followers to gain more information about them.  

\vspace{1mm}
Account details requests are limited to 15 requests per 15 minute window, and each request returns details for up to 100 accounts.  Similarly, ID list requests are limited to 180 requests per 15 minute window, and each request returns up to 5000 account IDs.  Hence, it is relatively easy to scrape the first-order honeypot account follower IDs and details without exceeding the rate limit, but collecting details for the second-order followers is a bottleneck. Since the number of nodes to collect information for can explode substantially even at the second-order, we limit collection to <100K friends and followers for each of the given follower of the honeypot account.  
We determine periodicity values empirically using back-of-the-envelope calculations.  While this data could be collected slowly using a single Twitter API key, we speed up the process by using multiple keys and cycling keys upon resource exhaustion.


\subsubsection{Preliminaries}

In the remainder of our work, we conduct analysis on two types of networks: the \emph{ego network} and \emph{boomerang network}.  

\vspace{1mm}
\noindent \textbf{Ego network:} An ego network (or egonet) traditionally consists of a central node called the ego, as well as the neighboring nodes and the relationships (edges) between them.  Egonets can essentially be considered as a local graphical representation of a node within the context of the broader, global graph and depict how the surrounding nodes are connected.  For our purposes, we examine \emph{per-service} egonets, where we consider the union of the individual egonets of both honeypot accounts per service.  Thus, in our case, each per-service egonet is actually comprised by 2 egos (the honeypot accounts), the union of both honeypots' neighboring nodes (the purchased, fake followers) and the relationships between them.  The per-service egonet representation allows us to both individually study the \emph{per-honeypot} egonets as well as any interactions between them.  That is, if the two honeypots for each service have distinct sets of neighboring nodes, then their per-honeypot egonets will also be distinct.  Conversely, if any nodes are neighbors of both honeypots, the associated per-honeypot egonets will be conjoined.  Various levels of overlap suggest differences with regards to how services reuse accounts to deliver fake links.

\vspace{1mm}
\noindent \textbf{Boomerang network:} Drawing conclusions from per-service egonet analysis can be deceiving in the sense that while it does give insights into the \emph{internal} relationships between the fake followers and honeypots, it does not consider the \emph{external} relationships formed by the fake followers.  As such, it is unable to give us a full perspective on the utilization of these fake followers.  In order to gain the requisite perspective, we conduct analysis of the proposed boomerang network.  We define the per-service boomerang network to be comprised of the per-service egonet \emph{in addition to} the out-links of the follower nodes -- the structure is reminiscent of a boomerang, in that it is comprised of the nodes ``1 step back and 1 step forward'' with respect to the honeypot account. Thus, the per-service boomerang network gives us an additional layer of information on top of the per-service egonet: connections to the other accounts followed by the honeypot's fake followers.  

\vspace{1mm}
We further use the \emph{density}, \emph{bipartite density}, \emph{transitivity} and \emph{reciprocity} metrics to summarize and describe network structure, and \emph{overlap coefficient} and \emph{multiple systems estimation} (MSE) to characterize network overlap.

\vspace{1mm}
\noindent \textbf{Density:} We define density as
\begin{displaymath}
\frac{\mathrm{\# edges}}{\mathrm{\# nodes} \; \cdot \; (\mathrm{\# nodes} \; - \; 1)}
\end{displaymath}
\noindent Density represents the fraction of existing to possible total edges, with density 1 indicating a complete graph.  

\vspace{1mm}
\noindent \textbf{Bipartite density:} We define bip. density between sets $\mathcal{A}$ and $\mathcal{B}$ as 
\begin{displaymath}
\frac{\mathrm{\# edges \; between} \; \mathcal{A} \; \mathrm{and} \; \mathcal{B}}{(\mathrm{\# nodes \; in} \; \mathcal{A}) \; \cdot \; (\mathrm{\# nodes \; in \;} \mathcal{B})}
\end{displaymath}
\noindent Bipartite density captures the fraction of existing to possible edges between two sets of nodes, with bipartite density 1 indicating a complete bipartite graph. 

\vspace{1mm}
\noindent \textbf{Transitivity:} We define transitivity as 
\begin{displaymath}
\frac{3 \; \cdot \; \mathrm{\# triangles}}{\mathrm{\# connected \; triples}}
\end{displaymath}
\noindent Transitivity denotes the degree of triadic closure, with transitivity 1 indicating that all connected triples of nodes are also triangles. 

\vspace{1mm}
\noindent \textbf{Reciprocity:} We define reciprocity as
\begin{displaymath}
\frac{\mathrm{\# bidirectional \; edges}}{\mathrm{\# edges}}
\end{displaymath}
\noindent Reciprocity conveys the relative frequency of bidirectional edges, with reciprocity 1 indicating that all edges are bidirectional.  

\vspace{1mm}
\noindent \textbf{Overlap coefficient:} We define overlap coef. between $\mathcal{A}$ and $\mathcal{B}$ as
\begin{displaymath}
\frac{|\mathcal{A} \; \cap \; \mathcal{B}|}{min(|\mathcal{A}|,|\mathcal{B}|)}
\end{displaymath}
\noindent Overlap coefficient indicates the proportion of members that overlap between sets, with overlap coefficient 1 indicating that $\mathcal{A} \subseteq \mathcal{B}$ or $\mathcal{B} \subseteq \mathcal{A}$ and 0 indicating $\mathcal{A} \cap \mathcal{B} = \emptyset$.

\vspace{1mm}
\noindent \textbf{Multiple systems estimation:} We use MSE to estimate population size from two randomly sampled sets $\mathcal{A}$ and $\mathcal{B}$ as
\begin{displaymath}
\frac{|\mathcal{A}| \; \cdot \; |\mathcal{B}|}{|\mathcal{A} \; \cap \mathcal{B}|}
\end{displaymath}
\noindent Intuitively, if $\mathcal{A}$ and $\mathcal{B}$ have low overlap, the total population size is much larger than if they have high overlap.

\vspace{1mm}
Upon shifting our discussion to account attributes distributions, we use \emph{entropy} as a means to capture distributional skew.

\vspace{1mm}
\noindent \textbf{Entropy:} We define entropy for a distribution $X$ with $n$ outcomes $(x_1 \ldots x_n)$ as 
\begin{displaymath}
-\sum\limits_{i = 1}^n P(x_i) \cdot \log_2 P(x_i)
\end{displaymath}
\noindent Entropy measures the unpredictability of a distribution in bits of information, with entropy of 0 bits indicating concentration of 100\% probability on a single outcome, and entropy of $\log_2 n$ bits indicating uniform distribution of probability between $n$ outcomes.
\subsection{Network Observations}

\begin{figure}[t!]
 \centering
 \begin{subfigure}[t]{0.22\textwidth}
 \includegraphics[width=\textwidth]{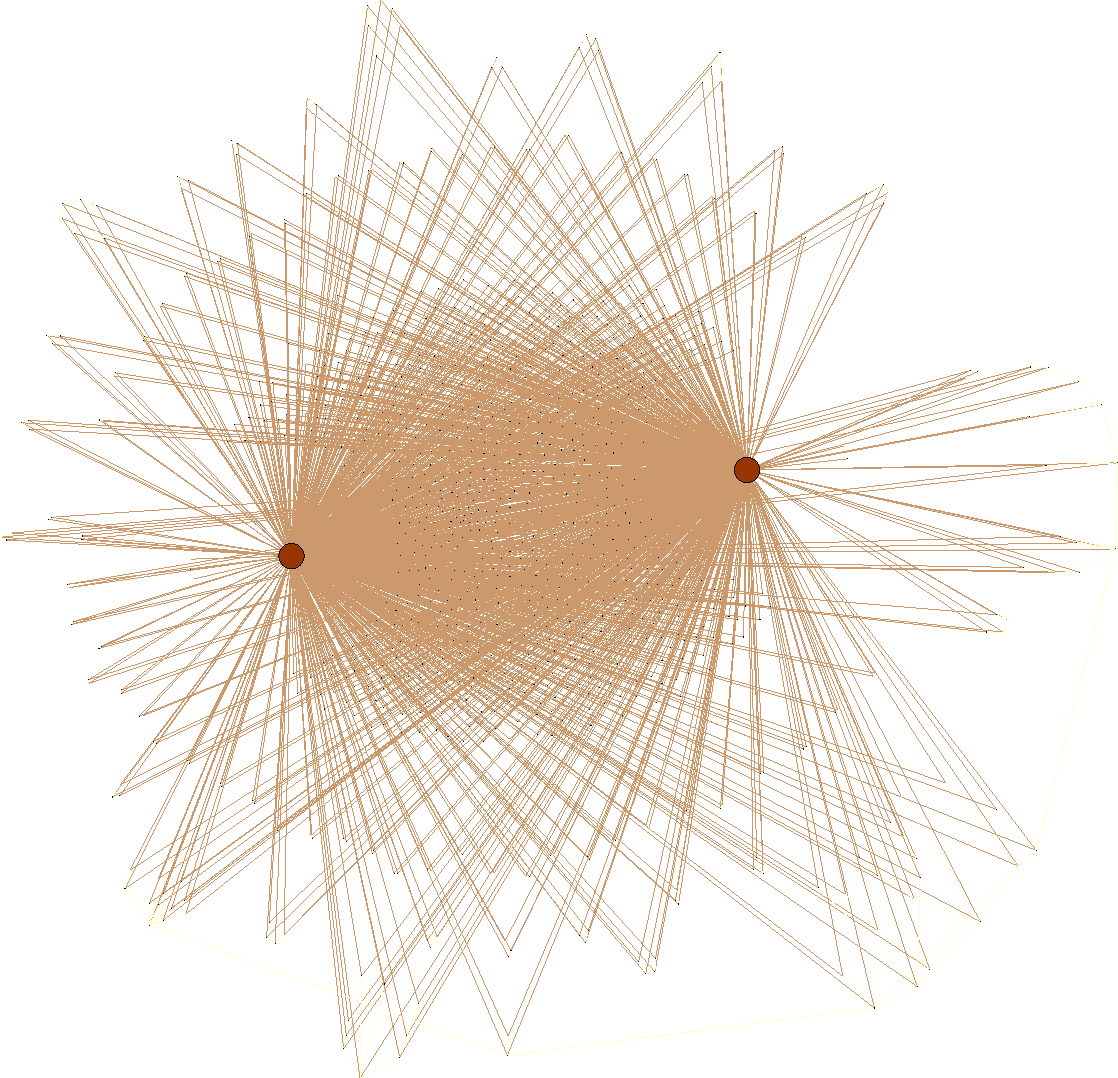}
 \caption[\fastfollowerz egonet]{\scriptsize\fastfollowerz}
 \label{fig:fastfollowerz_egonet}
 \end{subfigure} \hfill
 \begin{subfigure}[t]{0.22\textwidth}
 \includegraphics[width=\textwidth]{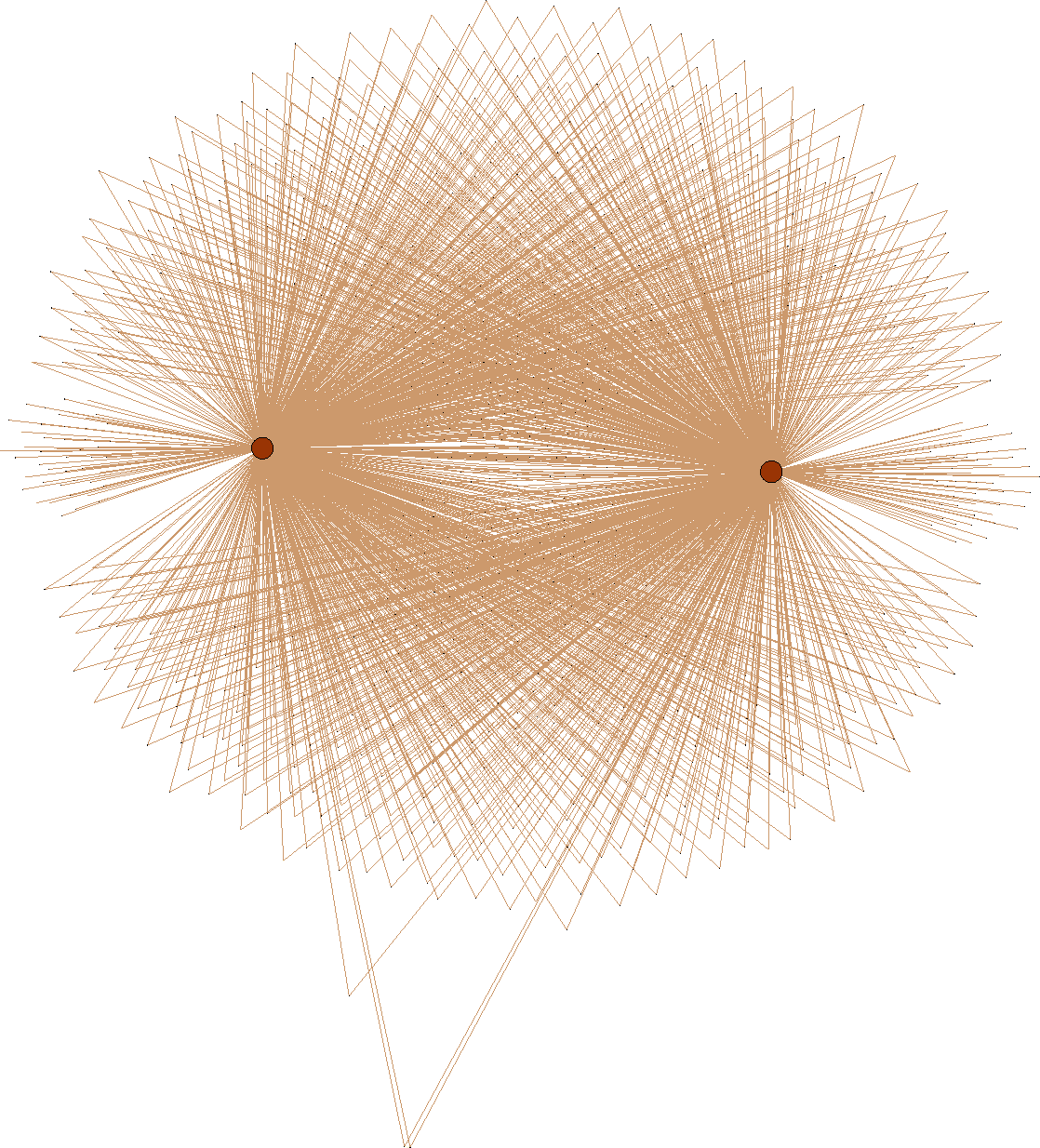}
 \caption[\intertwitter egonet]{\scriptsize\intertwitter}
 \label{fig:intertwitter_egonet}
 \end{subfigure} \hfill
 \begin{subfigure}[t]{0.22\textwidth}
 \includegraphics[width=\textwidth]{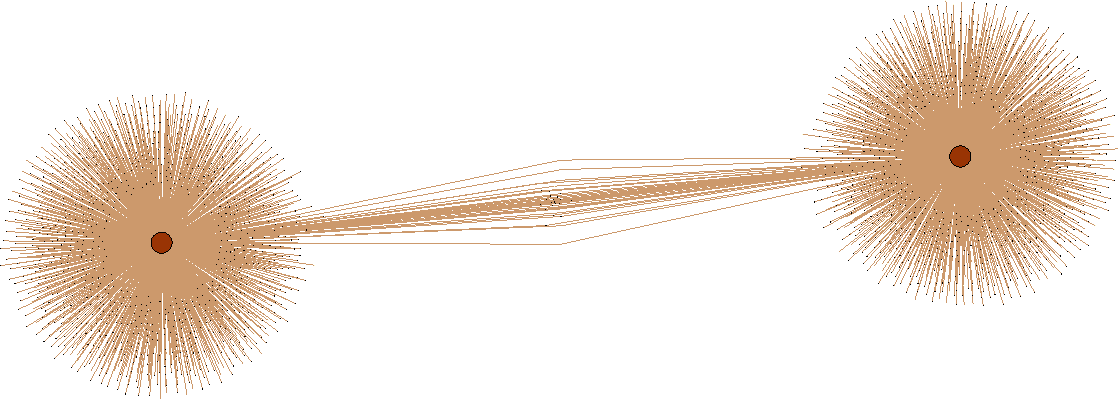}
 \caption[\devumi egonet]{\scriptsize\devumi}
 \label{fig:devumi_egonet}
 \end{subfigure} \hfill
  \begin{subfigure}[t]{0.22\textwidth}
 \includegraphics[width=\textwidth]{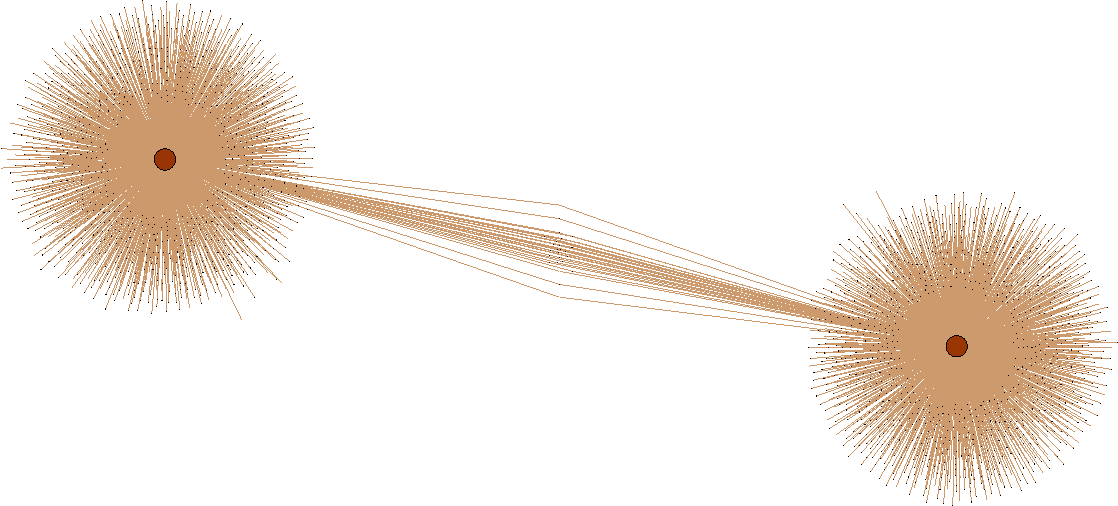}
 \caption[\twitterboost egonet]{\scriptsize\twitterboost}
 \label{fig:twitterboost_egonet}
 \end{subfigure}
 \medskip
 \hrule
 \medskip
 \begin{subfigure}[t]{0.22\textwidth}
 \includegraphics[width=\textwidth]{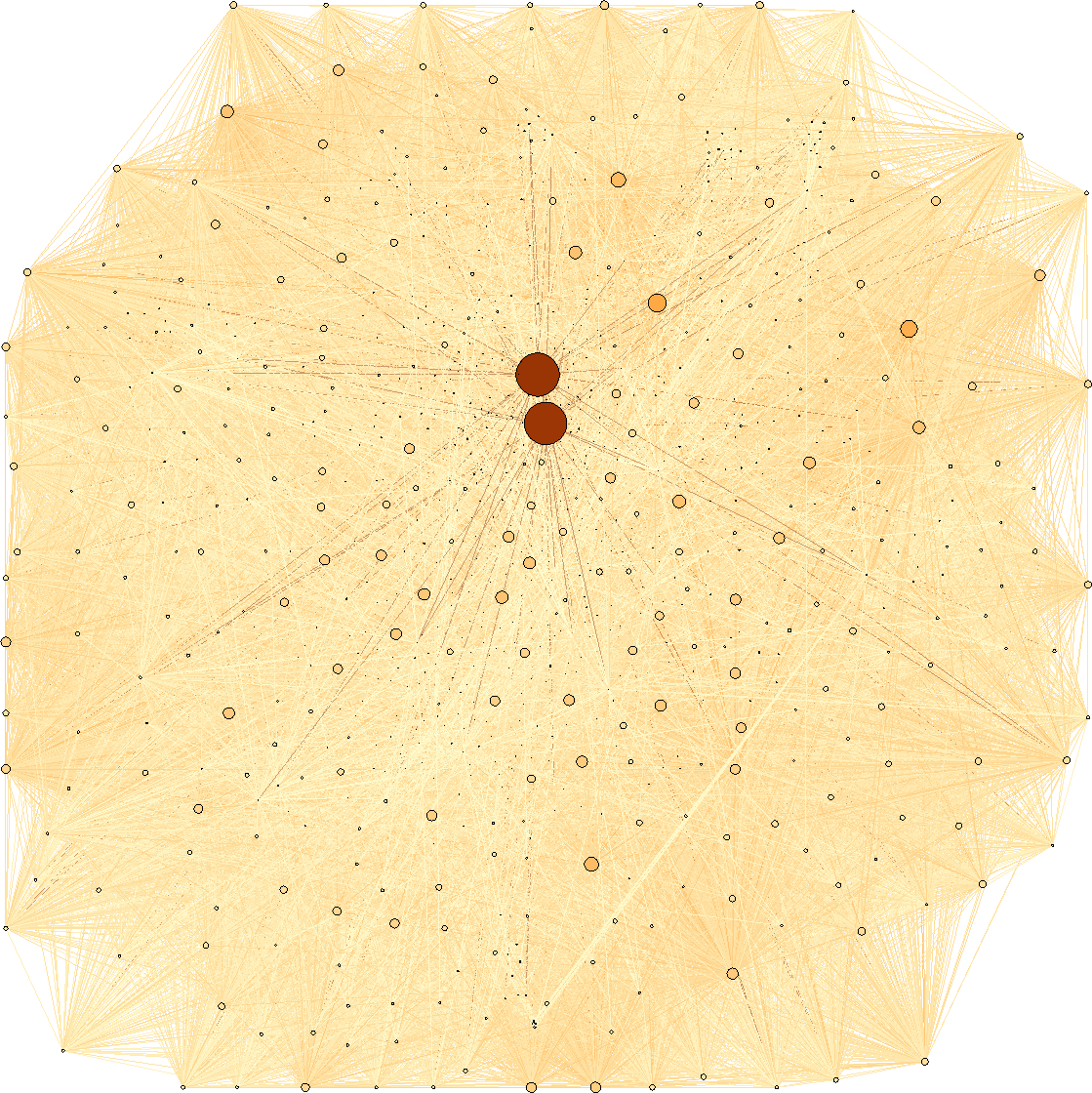}
 \caption[\plusfollower egonet]{\scriptsize\plusfollower}
 \label{fig:plusfollower_egonet}
 \end{subfigure} \hfill
 \begin{subfigure}[t]{0.22\textwidth}
 \includegraphics[width=\textwidth]{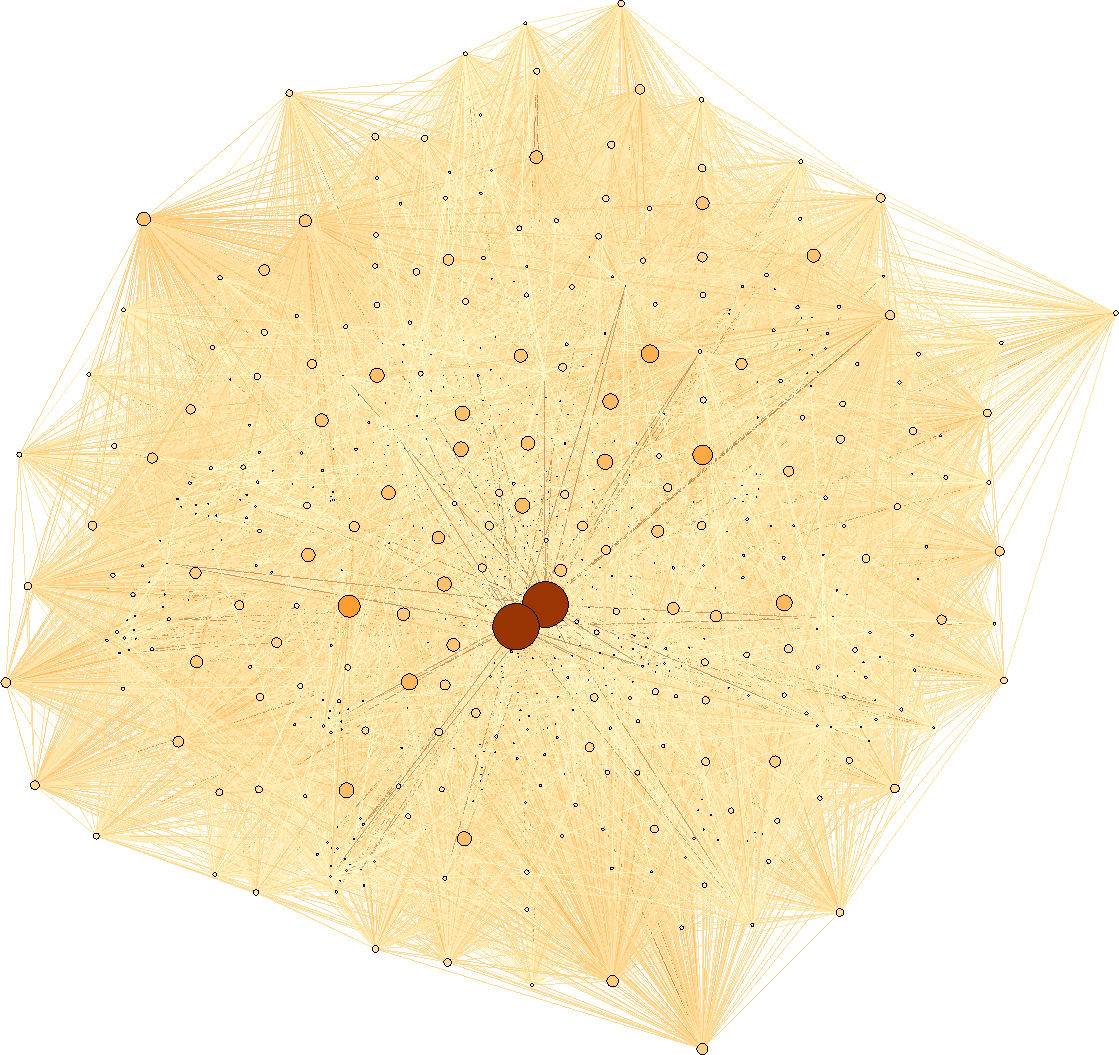}
 \caption[\newfollow egonet]{\scriptsize\newfollow}
 \label{fig:newfollow_egonet}
 \end{subfigure} \hfill
 \begin{subfigure}[t]{0.22\textwidth}
 \includegraphics[width=\textwidth]{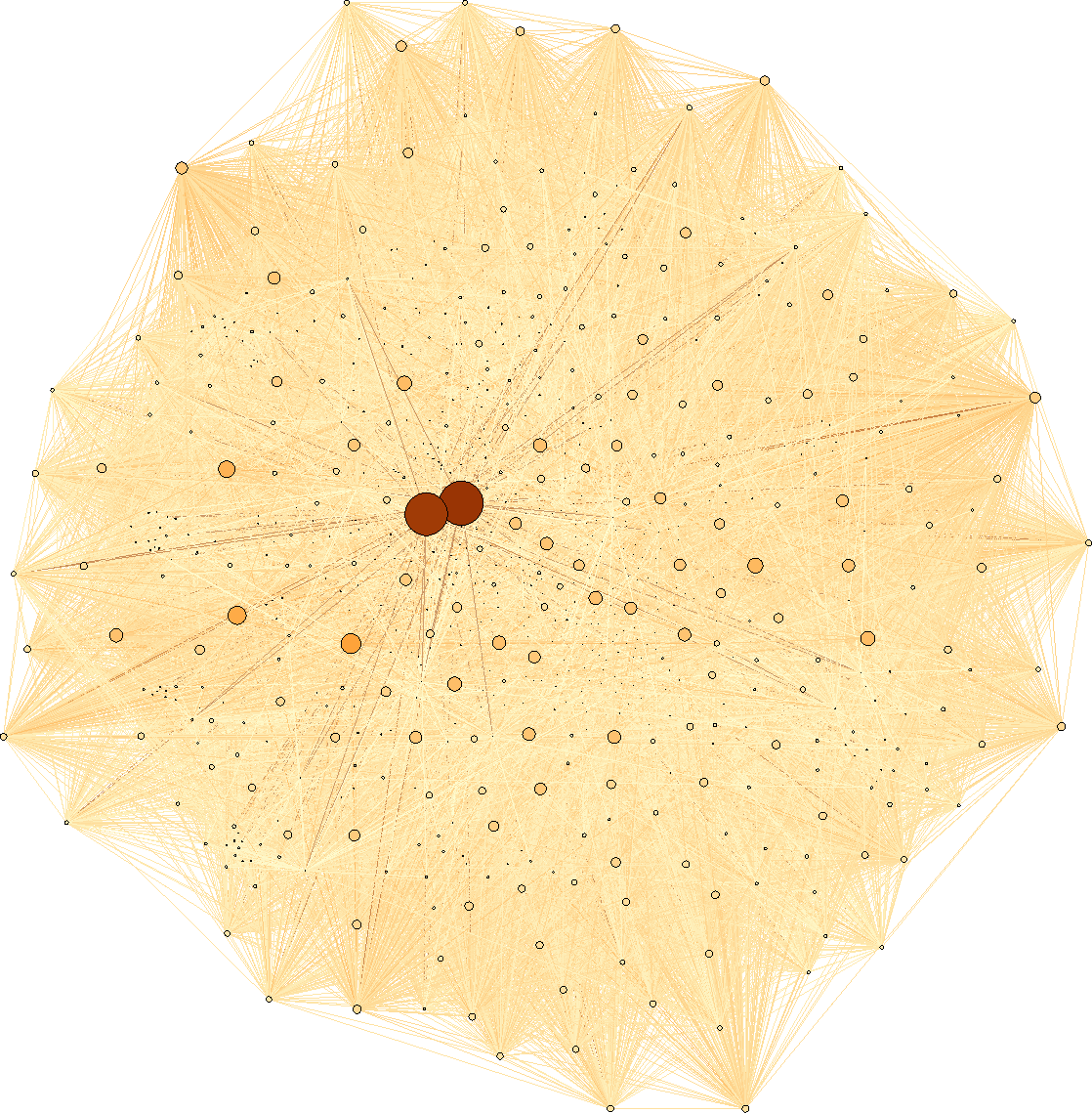}
 \caption[\hitfollow egonet]{\scriptsize\hitfollow}
 \label{fig:hitfollow_egonet}
 \end{subfigure} \hfill
 \begin{subfigure}[t]{0.22\textwidth}
 \includegraphics[width=\textwidth]{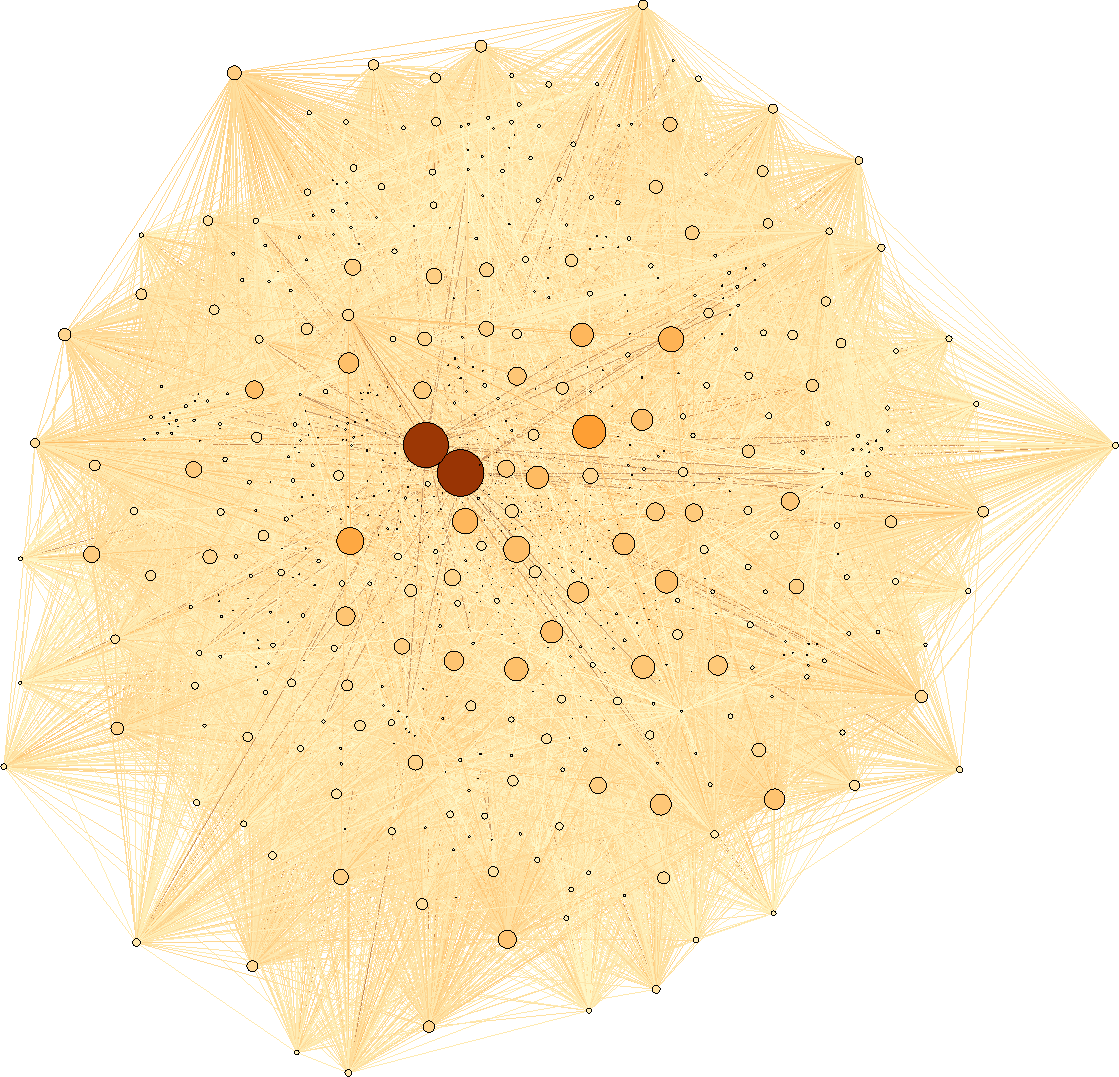}
 \caption[\bigfolo egonet]{\scriptsize\bigfolo}
 \label{fig:bigfolo_egonet}
 \end{subfigure} 
 
 \caption[Egonet differences in freemium and premium fraud]{\textbf{Premium fraudsters (top) form overlapping stars whereas freemium ones (bottom) form dense, near-cliques.}  Subplots show per-service egonets with honeypots in dark-red -- darker color and larger size indicates higher in-degree.}  
 \label{fig:egonets}
\end{figure}

We first focus on studying the local network properties of fraudulent accounts.  Targeting oddities in network connectivity is a central theme in many link fraud detection approaches, as the mission constraints of delivering fake links to customers necessarily affects graph structure.  But what are these changes?  In this section, we leverage social network analysis tools to characterize effects of fraud on the surrounding network structure, and show the similarities and differences between premium and freemium fraud.  We detail analyses on two types of induced subgraphs: the ego network and more expansive boomerang network.  

\subsubsection{Ego Network Patterns}

Figure \ref{fig:egonets} shows the per-service egonets for each of the 8 providers, with increased node size and darkness corresponding to higher in-degree. 
The honeypots (egos) are the two large and dark orange colored nodes in each subfigure. Cursory analysis reveals a notable difference in egonet network structure between freemium and premium providers.  We see that the premium egonets (first row) have a star/bipartite structure: each honeypot node is the hub of a star, and the satellite nodes overlap and are disconnected.  Conversely, freemium egonets have denser, near-clique type structure which suggests denser connectivity between the neighboring nodes.  

The statistics for premium service egonets in Table \ref{tbl:egonet_stats} (top) further lend credence to the visual differences we observe from Figure \ref{fig:egonets}, giving us the following insight:

\begin{insight}[Egonet Sparsity]
\label{ins:star}
Premium fake followers rarely follow each other, resulting in sparse egonet structure. Freemium fake followers have dense egonet structure.
\end{insight}

This is substantiated by the low density and node to edge ratios across premium providers.  Of these, \fastfollowerz and \intertwitter have an order of magnitude greater density than \devumi and \twitterboost.  This is substantiated by the 1:2 node to edge ratio in the former 2 providers as compared to the near 1:1 ratios of the latter 2.  \fastfollowerz and \intertwitter also have marginally higher transitivity values compared to the 0 transitivity of \devumi and \twitterboost, indicating that the former 2 have few triangles between the fake follower nodes whereas the latter 2 have none.  We also observe no reciprocal links in these providers, indicating only one-way relationships.

Conversely, the freemium statistics in Table \ref{tbl:egonet_stats} (bottom) support that freemium fake followers have dense egonet structures. Freemium providers are an order of magnitude denser than the densest premium egonets -- all 4 providers have 6-7\% density.  While not shown in interest of space, the per-honeypot egonets were each found to have an even higher 11-14\% density individually.  The 1:50 node to edge ratios substantiate this high density.  We also notice that transitivity values are much higher for freemium providers, suggesting that an unusually high 28-30\% of wedges are also triangles.  Given that density and transitivity are equal in random graphs, the freemium egonets do not appear to be random, but are likely composed of dense subregions which are themselves sparsely connected.  The link structure reflects how freemium providers trade follows between accounts (random partitions, biased selection, account similarity, etc.)  Furthermore, all 4 providers have similar, high reciprocity of 40-42\% suggesting frequent ``follow-back'' behavior.

%

\vspace{1mm}
\noindent \textbf{Rationale:} 
The freemium services accumulates a pool of free accounts, and hence trading follows enables each free user to gain some followers.  As a result, 
such behavior creates a denser subgraph, but are also used by providers to deliver the follower demands of paid customers and turn a profit.  Comparatively, premium providers are unable to use free users' accounts and must create fake accounts. 

These insights pose an interesting question: as we expect fraudsters to act in a manner that maximizes profit, \emph{what motivates the differences in structure between freemium and premium providers?}  We propose an answer: If we consider that each account has a budget of edges it can create without being suspended, it seems that premium providers greatly underutilize accounts compared to freemium ones.  This is because for fraudsters, delivering more links while avoiding suspension is strictly better as it means that they can either serve more customers or artificially inflate their own popularity.  

\begin{table}[t!]
\renewcommand{\arraystretch}{0.9}
\setlength\tabcolsep{2pt}
\centering
\caption[Egonet summary statistics]{Egonet summary statistics.}
\label{tbl:egonet_stats}
\begin{tabular}{c@{}llllll@{}}
\toprule
&\textbf{Service}  & \textbf{\# Nodes} & \textbf{\# Edges} & \textbf{Density} & \textbf{Transitivity} & \textbf{Reciprocity} \\ \midrule
\multirow{4}{*}{\rotatebox[origin=c]{90}{\underline{Premium}}\quad} & \fastfollowerz & 1,066              & 2,289              & .002             & .001                  & .000                    \\
&\intertwitter  & 1,051              & 2,003              & .002             & .00006                     & .000                    \\
&\devumi        & 2,681              & 2,712              & .0003            & .000                     & .000                    \\
&\twitterboost   & 2,680              & 2,711              & .0004            & .000                     & .000                    \\ \midrule
\multirow{4}{*}{\rotatebox[origin=c]{90}{\underline{Freemium}}\quad} & \plusfollower & 920               & 51,868             & .061             & .288                  & .411                 \\
&\newfollow    & 755               & 37,052             & .065             & .294                  & .408                 \\
&\hitfollow    & 782               & 41,879             & .068             & .305                  & .416                 \\
&\bigfolo      & 749               & 36,043             & .064             & .294                  & .413                 \\ \bottomrule
\end{tabular}
\end{table}

\subsubsection{Boomerang Network Patterns}

\begin{figure}[t!]
 \centering
 \begin{subfigure}[t]{0.45\textwidth}
 \includegraphics[width=\textwidth]{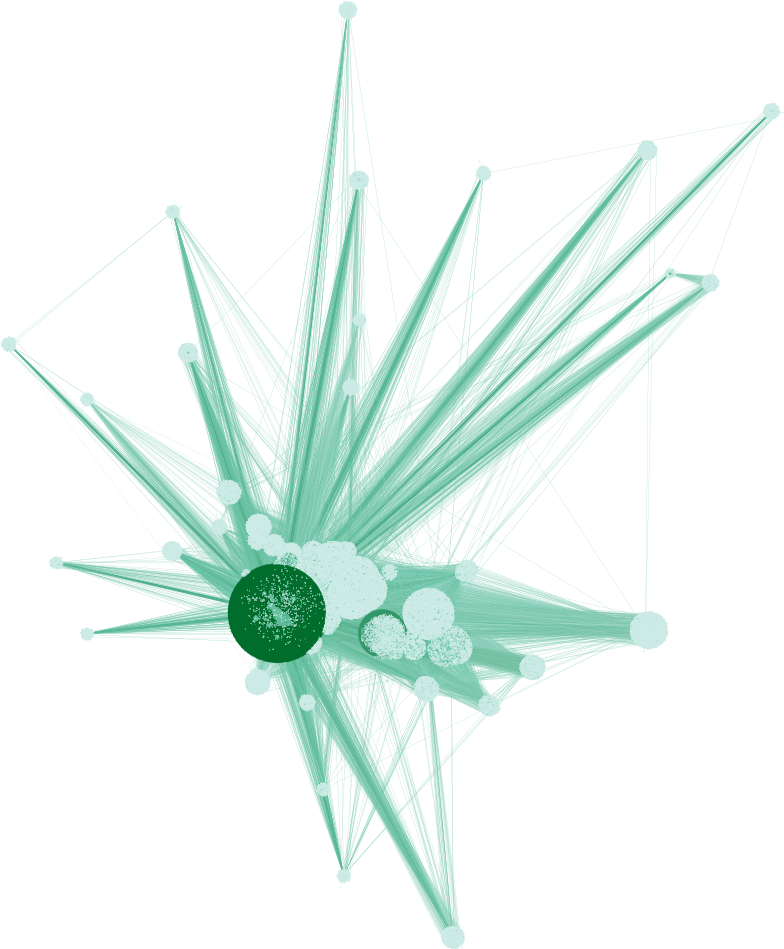}
 \caption[\bigfolo boomerang network]{\bigfolo (fre.)}
 \label{fig:bigfolo_boomerang}
 \end{subfigure}
 \hfill
 \begin{subfigure}[t]{0.45\textwidth}
 \includegraphics[width=\textwidth]{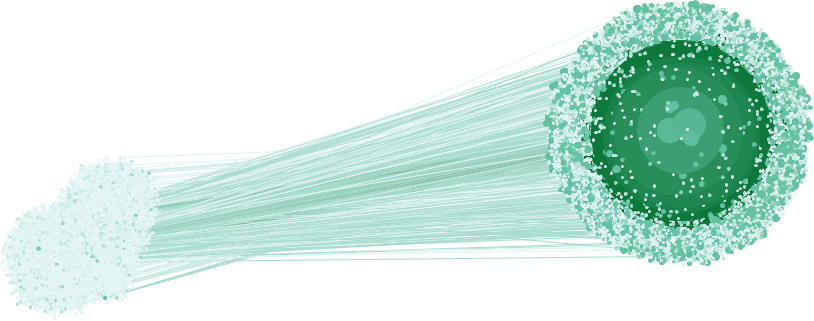}
 \caption[\twitterboost boomerang network]{\twitterboost (pre.)}
 \label{fig:twitterboost_boomerang}
 \end{subfigure}
 \caption[Boomerang network differences in freemium and premium fraud]{\textbf{Freemium followers have dense internal and sparse external connectivity (top), and vice versa for premium followers (bottom).} Subplots show boomerang networks, with darker node color and larger size indicating higher in-degree.}
 \label{fig:boomerangs}
\end{figure}

Figure \ref{fig:boomerangs} shows 2 boomerang networks, one for \bigfolo and \twitterboost, each representative of a different fraud strategy.  
Again, honeypot accounts are amongst the large, dark nodes with high in-degree, and the lighter, smaller nodes are fake followers or their friends.  Note that the layout clusters nodes based on similar linkage, so groups of nodes visually close share connectivity properties.  As with egonets, we again see a stark contrast in the boomerang structure of these two providers.  Figure \ref{fig:bigfolo_boomerang} shows the dense internal connectivity of \bigfolo's fake followers (as we saw in Figure \ref{fig:bigfolo_egonet}), in conjunction with the sparser and less compact external connectivity to friends.  Conversely, Figure \ref{fig:twitterboost_egonet} shows sparse internal connectivity between \twitterboost's fake followers on the left, but dense near-bipartite external connectivity to the customers (including honeypots) on the right. 

Table \ref{tbl:boomerang_stats} (top) gives summary statistics about premium boomerang networks, which substantiate the following:

\begin{insight}[Boomerang Density]
\label{ins:highReuse}
Premium fake followers are frequently reused to follow customers, resulting in dense external connectivity in the boomerang network. Freemium fake followers are less reused to follow customers, and hence have sparse external connectivity.
\end{insight}

Interestingly, we see that the relative values of these statistics are inverted for the boomerang networks from the egonets -- unlike for egonets where the density metric was an order of magnitude higher for freemium providers, the bipartite density in boomerang networks is instead an order of magnitude higher for the premium providers.
 Note that the premium providers' bipartite density indicates that nearly 1-2\% (a huge amount) of all possible edges between the fake followers and their combined set of friends exists.  The node to edge ratios are also much higher for premium providers -- \fastfollowerz and \intertwitter are 1:14, and \devumi and \twitterboost are roughly 1:37 compared to only 1:2 for the freemium providers.

The freemium boomerang network statistics in Table \ref{tbl:boomerang_stats} (bottom) again establishes the second part of the insight. This is further substantiated by the observation that freemium providers have an order of magnitude lower bipartite density than premium ones. We also observe that freemium boomerang networks have higher number of nodes than the premium counterpart. This is intuitive as freemium followers are otherwise genuine accounts, they have an expansive set of true friends, whereas premium fake followers are all synthetic accounts.



\begin{table}[t!]
\renewcommand{\arraystretch}{0.9}
\setlength\tabcolsep{5pt}
\centering
\caption[Boomerang network summary statistics]{Boomerang network summary statistics.}
\label{tbl:boomerang_stats}
\begin{tabular}{c@{}llll@{}}
\toprule
&\textbf{Service} & \textbf{\# Nodes} & \textbf{\# Edges} & \textbf{Bip. Density} \\ \midrule
\multirow{4}{*}{\rotatebox[origin=c]{90}{\underline{Premium}}\quad} &\fastfollowerz   & 40,486            & 491,458           & .012                 \\
&\intertwitter    & 176,921           & 2,383,251         & .013                   \\
&\devumi          & 67,893            & 2,495,586         & .014                  \\
&\twitterboost    & 68,297            & 2,474,759         & .014                 \\ \midrule
\multirow{4}{*}{\rotatebox[origin=c]{90}{\underline{Freemium}}\quad}& \plusfollower    & 646,901           & 1,352,253         & .002                  \\
&\newfollow       & 616,824           & 1,221,574         & .003                  \\
&\hitfollow       & 558,100           & 1,172,248         & .003                 \\
&\bigfolo         & 574,823           & 1,157,672         & .003                  \\ \bottomrule
\end{tabular}
\end{table}


\subsubsection{Network Overlap Patterns}

In our analysis thus far, we noticed that various providers have different levels of evident overlap in the fake followers they deliver between their 2 honeypots.  How extensive is this overlap?  Do these providers reuse accounts in the same ways?  Furthermore, is there any overlap between the followers across providers? Here, we shed light on these questions.

\paragraph{Intra-Network Patterns}

First, we study \emph{intra-network overlap}, describing overlap between the fake follower nodes within each service.  Table \ref{tbl:intranetoverlap} shows the overlap coefficients between the honeypot followers for each service.  Assuming the followers for each honeypot are randomly sampled from the service's account pool, we additionally compute the estimated total number of fake accounts currently in the fraud provider's hands using MSE.

\begin{table}[t!]
\renewcommand{\arraystretch}{0.9}
\centering
\caption[Fraudster account reuse habits]{Fraud providers have varying account reuse habits.}
\label{tbl:intranetoverlap}
\begin{tabular}{c@{}llll@{}}
\toprule
&\textbf{Service} & \textbf{\# Nodes} & \textbf{Overlap} & \textbf{Est. Pool \# Nodes} \\ \midrule
\multirow{4}{*}{\rotatebox[origin=c]{90}{\underline{Premium}}\quad} & \fastfollowerz   & 1,064             & .996             & 1,064                       \\
&\intertwitter    & 1,049             & .953             & 1,051                       \\
&\devumi          & 2,679             & .024             & 55,719                      \\
&\twitterboost    & 26,78             & .024             & 55,677                      \\ \midrule
\multirow{4}{*}{\rotatebox[origin=c]{90}{\underline{Freemium}}\quad} & \plusfollower    & 918               & .815             & 954                         \\
&\newfollow       & 753               & .765             & 798                         \\
&\hitfollow       & 780               & .802             & 814                         \\
&\bigfolo         & 747               & .774             & 791                         \\ \bottomrule
\end{tabular}
\end{table}
The various degrees of overlap and commensurate estimates of pool size suggest the following insight:

\begin{insight}[Varying Delivery Structure]
\label{ins:variableDelivery}
Service providers have varying methods for account reuse in efforts to to distribute suspicion across their account pools.  
\end{insight}

We observe that the freemium providers tend to have a high, 0.8 overlap which results in an estimated pool size slightly larger than either of the two sets of honeypot followers.  However, the premium providers have an interesting split which reveals that \fastfollowerz and \intertwitter have very high, near 1.0 overlap, resulting in the pool size being roughly equal to each set of followers.  This indicates that the pool is reused almost exactly for multiple customers.  Conversely, \devumi and \twitterboost have near 0 overlap.  As a result, we estimate that the pool size is quite large, containing over 55K total fake accounts. 

While we cannot be certain without further investigation, these providers likely have different means of selecting and shifting the pool of active fake followers.  For example, the pools used in \fastfollowerz and \intertwitter may cycle between a number of different ``sub-pools'' based on time, customer account features, or random choice.  
Conversely, the evidently much larger estimated pool size for \devumi and \twitterboost suggests that they may each have a single, large fixed pool of usable accounts from which followers are sampled regardless of other factors.

\paragraph{Inter-Network Patterns}
Thus far, we have established that providers reuse multiple follower accounts across customers in order to turn a better profit.  But how far does this reuse go?  Are any accounts responsible for delivering fake links to customers from different providers?  To answer these questions, we study the pairwise \emph{inter-network overlap} of followers between providers.  

Table \ref{tbl:internetoverlap} shows an $8 \times 8$ matrix with the pairwise overlap coefficients.  Given the number of nonzero entries, we draw the following surprising insight:

\begin{insight}[Collusion]
\label{ins:collusion}
Service providers seem to collaborate with and draw from each other to commit fraudulent actions.
\end{insight}

We notice that there is substantial overlap within the freemium and premium providers.  While \fastfollowerz and \intertwitter share no accounts with the other premium providers, \devumi and \twitterboost have a .07 overlap.  Comparatively, all 4 freemium providers have a large 0.6-0.7 overlap, indicating that most of their fake accounts are \emph{the same}.  Furthermore, the set of followers for freemium and premium providers have 0 overlap, substantiating that followers in freemium providers are otherwise real accounts whereas those in premium providers are synthetic.  

Nonzero overlap between providers is an interesting finding -- it is indicative of either a willingness to share follower accounts between fraud providers, or commonality in leaked or hijacked accounts.  Upon further inspection, we notice a number of suggestive findings:

\begin{compactitem}
\item Overlapping providers shared domain WHOIS protectors.
\item Overlapping premium providers use the same Yoast SEO plugin and stylesheets.
\item All freemium providers have two-column sites, advertised up to 30K followers, and priced from \textsterling9.99.
\item All fremium providers contained the line: \textit{``[service] is Not Affiliated With OR Endorsed By Twitter.com.''}
\end{compactitem}

\begin{table}[t!]
\renewcommand{\arraystretch}{0.9}
\setlength\tabcolsep{5pt}
\centering
\caption[Collusion between fraud providers]{Fraud providers share follower accounts.}
\label{tbl:internetoverlap}
\begin{tabular}{c@{}lllll|llll@{}}
               && \rotatebox{90}{\fastfollowerz} & \rotatebox{90}{\intertwitter} & \rotatebox{90}{\devumi} & \rotatebox{90}{\twitterboost} & \rotatebox{90}{\plusfollower} & \rotatebox{90}{\newfollow} & \rotatebox{90}{\hitfollow} & \rotatebox{90}{\bigfolo} \\ \midrule
\multirow{4}{*}{\rotatebox[origin=c]{90}{\underline{Premium}}\quad} & \fastfollowerz & {1.0}            & 0           & 0     & 0           & 0           & 0      & 0        & 0      \\
&\intertwitter  & 0            & {1.0}            & 0     & 0           & 0           & 0      & 0        & 0      \\
&\devumi        & 0            & 0           & {1.0}       & {.07}            & 0           & 0      & 0        & 0      \\
&\twitterboost  & 0            & 0           & {.07}     & {1.0}             & 0           & 0      & 0        & 0      \\ \midrule
\multirow{4}{*}{\rotatebox[origin=c]{90}{\underline{Freemium}}\quad} & \plusfollower  & 0            & 0           & 0     & 0           & {1.0}             & {.65}      & {.69}        & {.64}      \\
&\newfollow       & 0            & 0           & 0     & 0           & {.65}           & {1.0}        & {.64}        & {.63}      \\
&\hitfollow     & 0            & 0           & 0     & 0           & {.69}           & {.64}      & {1.0}          & {.63}      \\
&\bigfolo       & 0            & 0           & 0     & 0           & {.64}           & {.64}      & {.63}        & {1.0}        \\ \bottomrule
\end{tabular}
\end{table}

\subsection{Attribute Observations}

\begin{table*}[t]
\renewcommand{\arraystretch}{0.9}
\setlength\tabcolsep{2pt}
\scriptsize
\centering
\caption[Varying entropy over fraudster account attributes]{Per-service entropy (in bits) over account attribute distributions.}
\label{tbl:attrib_entropy}
\scalebox{0.8}{
\begin{tabular}{c@{}llllllllllllll@{}}
\toprule
&\textbf{Service}      & \textbf{Created (year)} & \textbf{Def. Prof.} & \textbf{Def. Prof. Image} & \textbf{\# Favorites} & \textbf{\# Followers} & \textbf{\# Friends} & \textbf{\# Lists} & \textbf{\# Statuses} & \textbf{Geolocation} & \textbf{Lang.} & \textbf{Protected} & \textbf{UTC} & \textbf{Verified} \\ \midrule
\multirow{4}{*}{\rotatebox[origin=c]{90}{\underline{Premium}}\quad} & \fastfollowerz        & 1.37                 & .63                       & .01                              & 3.65                       & 2.73                      & 2.73                    & 2.99                   & 3.8                      & .00                   & .06           & .00                & 1.04                 & .00               \\
&\intertwitter         & 2.99                 & .82                       & .94                              & 4.04                       & 3.54                      & 2.63                    & 2.53                   & 4.31                     & .67                   & 2.55          & .56                & 1.97                 & .18               \\
&\devumi               & 1.13                 & .97                       & .02                              & 1.05                       & 1.54                      & 1.17                    & 2.49                   & 1.18                     & .00                   & .00           & .00                & 1.42                 & .00               \\
&\twitterboost         & 1.13                 & .97                       & .03                              & 1.05                       & 1.56                      & 1.16                    & 2.51                   & 1.15                     & .00                   & .00           & .00                & 1.41                 & .00               \\ \midrule
\multirow{4}{*}{\rotatebox[origin=c]{90}{\underline{Freemium}}\quad} & \plusfollower         & 1.82                 & .93                       & .73                              & 4.18                       & 3.76                      & 3.38                    & 2.73                   & 4.40                     & .54                   & 2.04          & .30                & 1.70                 & .00               \\
&\newfollow            & 1.68                 & .90                       & .75                              & 4.20                       & 3.70                      & 3.32                    & 2.64                   & 4.37                     & .55                   & 1.99          & .28                & 1.62                 & .00               \\
&\hitfollow            & 1.78                 & .93                       & .73                              & 4.14                       & 3.76                      & 3.32                    & 2.72                   & 4.37                     & .52                   & 2.01          & .30                & 1.70                 & .00               \\
&\bigfolo              & 1.88                 & .92                       & .75                              & 4.20                       & 3.74                      & 3.34                    & 2.72                   & 4.40                     & .56                   & 2.05          & .32                & 1.71                 & .00               \\ \midrule\midrule
&\textbf{Max Entropy:} & 3.46                 & 1.00                      & 1.00                             & 5.00                       & 5.00                      & 5.00                    & 5.00                   & 5.00                     & 1.00                  & 5.13          & 1.00               & 5.29                 & 1.00              \\ \bottomrule
\end{tabular}}
\end{table*}

In this section, we study the similarities and differences in account attributes of fake followers.  
Table \ref{tbl:attrib_entropy} shows per-service, per-attribute entropy in bits for a variety of user attributes.  The account attributes include creation year, default profile and profile image booleans, favorites count, followers count, friends count, lists count, statuses count, geolocation enabled boolean, language identifier, protected statuses boolean, UTC timezone, and a Twitter verification boolean which corresponds to high-profile, ``famous'' accounts.  
These attributes have varying outcome spaces.  Creation date has 11 possible years (2006-2016), since Twitter was founded in 2006.  Booleans have 2 possible outcomes (T,F).  We encountered 35 different language identifiers and 39 UTC timezone settings.  For count features, we logarithmically discretized the space into 32 bins from 1 to 1M to capture the wide range of activity levels.  For each service, we aggregate attribute values and compute the entropy over the outcomes.  The table shows the actual sample entropy in addition to the maximum possible (uniform) entropy.  As previously mentioned, lower entropy indicates high synchronicity between followers.  Note that a difference in entropy of 1 bit corresponds to twice the predictability.  

The most striking insight from Table \ref{tbl:attrib_entropy} is as follows:

\begin{insight}[Entropy Gap]
\label{ins:entropy}
Premium service providers deliver followers with low entropy, high regularity attributes, whereas freemium service providers have more attribute disparity.
\end{insight}

We notice that the premium providers have substantially lower entropy values in many attributes versus freemium providers, and even near 0 entropy in other attributes like geolocation.  We elaborate on the specific differences next.

\subsubsection{Account Creation}
\devumi, \twitterboost and \fastfollowerz have very low creation year entropy compared to freemium providers.  While both freemium and premium accounts tend to be created more recently (perhaps because of higher suspension rate in older accounts), premium providers have a heavy bias towards recently created accounts (>2014).  

\subsubsection{Profile Defaults}
\fastfollowerz has a much lower entropy than other providers in terms of default profile -- we found that >84\% of these accounts did not have a default profile, whereas default profiles are actually \emph{more common} than not in freemium accounts.  Surprisingly, \fastfollowerz, \devumi and \twitterboost also have near 0 entropy for profile image compared to the much higher entropy for freemium providers.  We find that premium followers almost always set a custom image, suggesting that the information was fabricated or stolen from real users. Conversely, default profile images are common for freemium service accounts -- this is intuitive, most real users do not fully customize their profiles.

\subsubsection{Action Counts}
\begin{figure*}[t!]
  \centering
  \begin{tabular}{cc} 
   \begin{subfigure}[t]{0.43\textwidth}
    \centering
    \includegraphics[width=\textwidth]{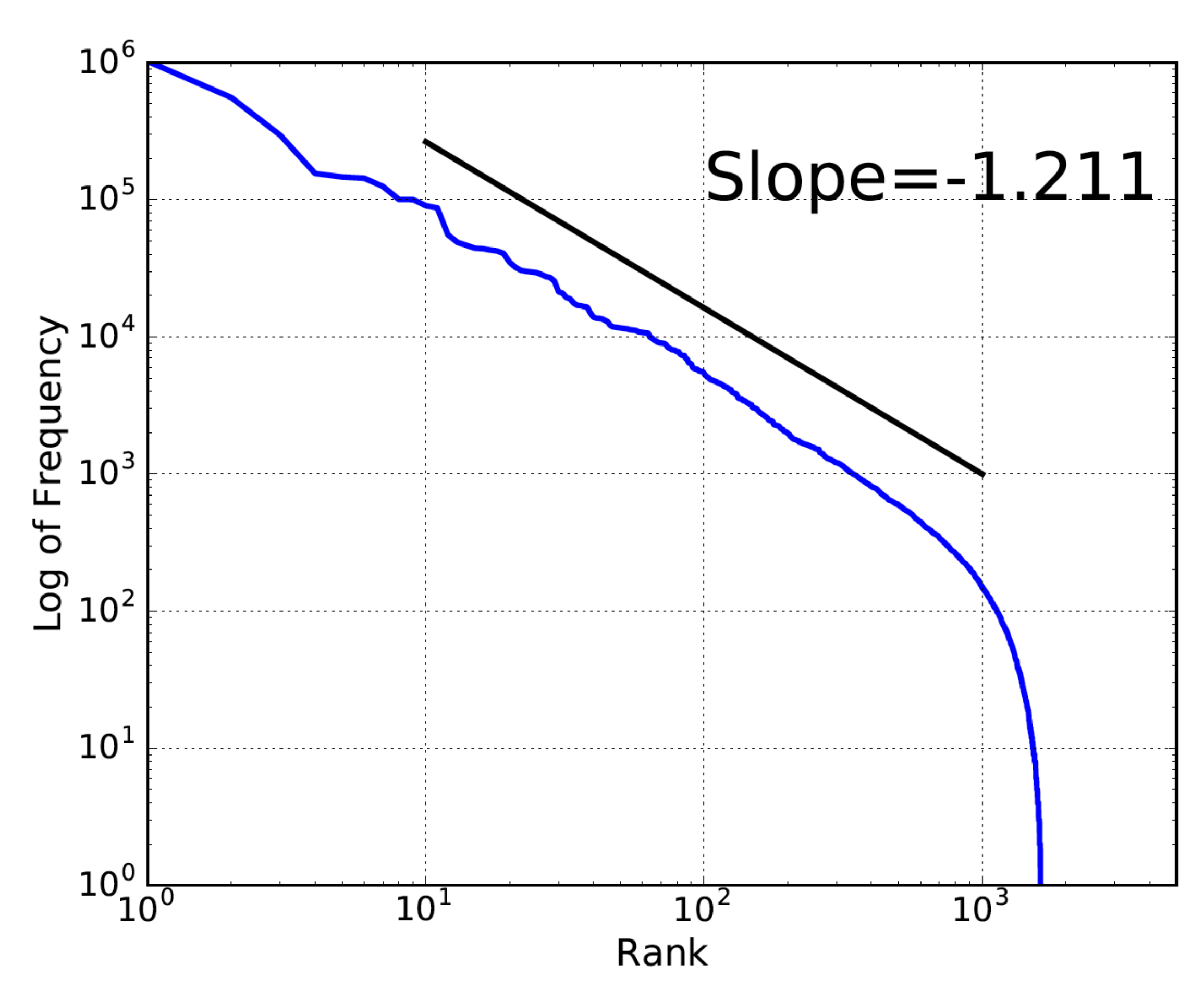}
    \caption[Genuine user follow counts]{Gen. follow count}
    \label{fig:rf_gen_followers}
   \end{subfigure}
  &
   \begin{subfigure}[t]{0.43\textwidth}
    \centering
    \includegraphics[width=\textwidth]{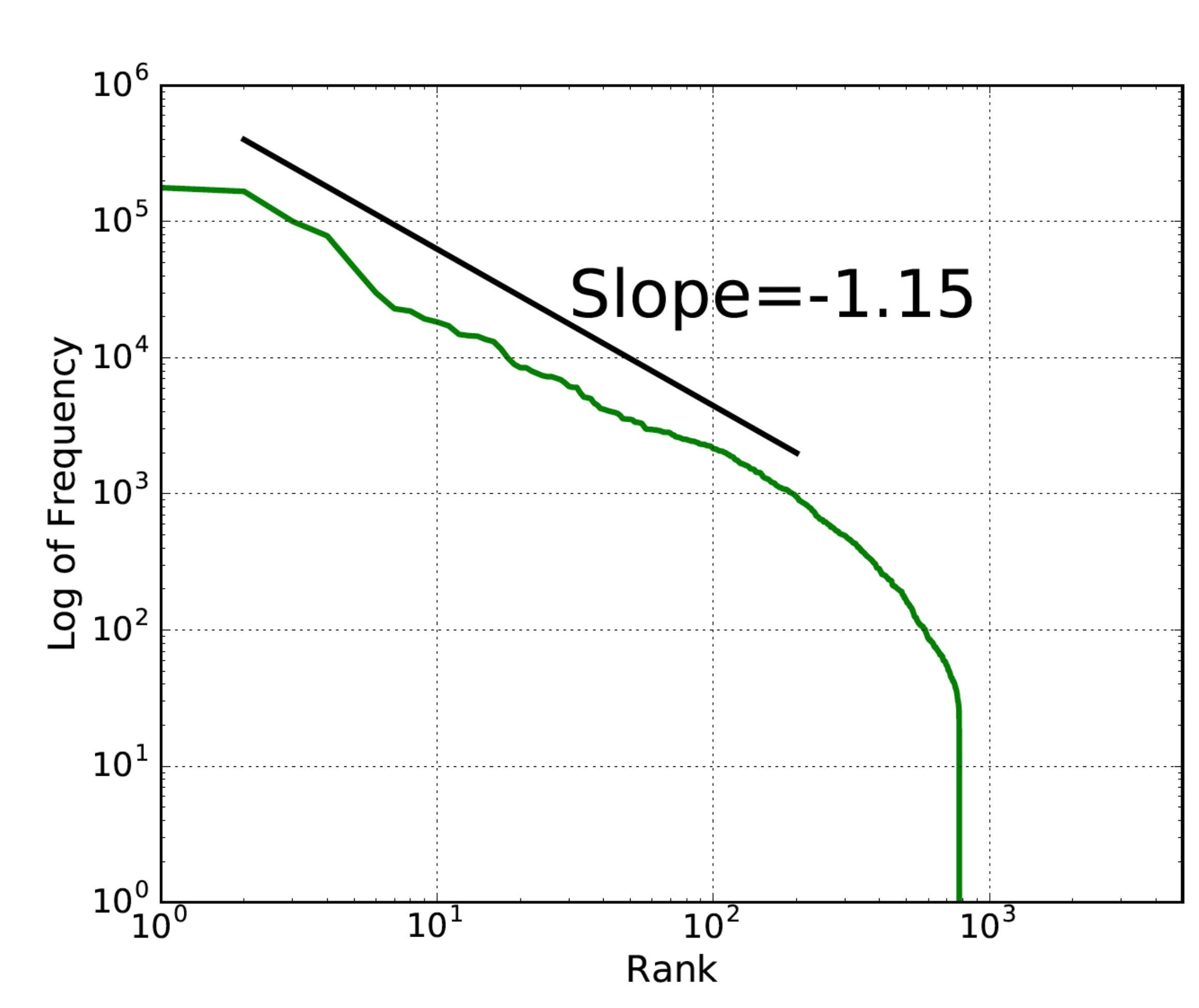}
    \caption[Freemium fraudster follow counts]{Fre. follow count}
    \label{fig:rf_free_followers}
   \end{subfigure} \\

   \begin{subfigure}[t]{0.43\textwidth}
     \centering
     \includegraphics[width=\textwidth]{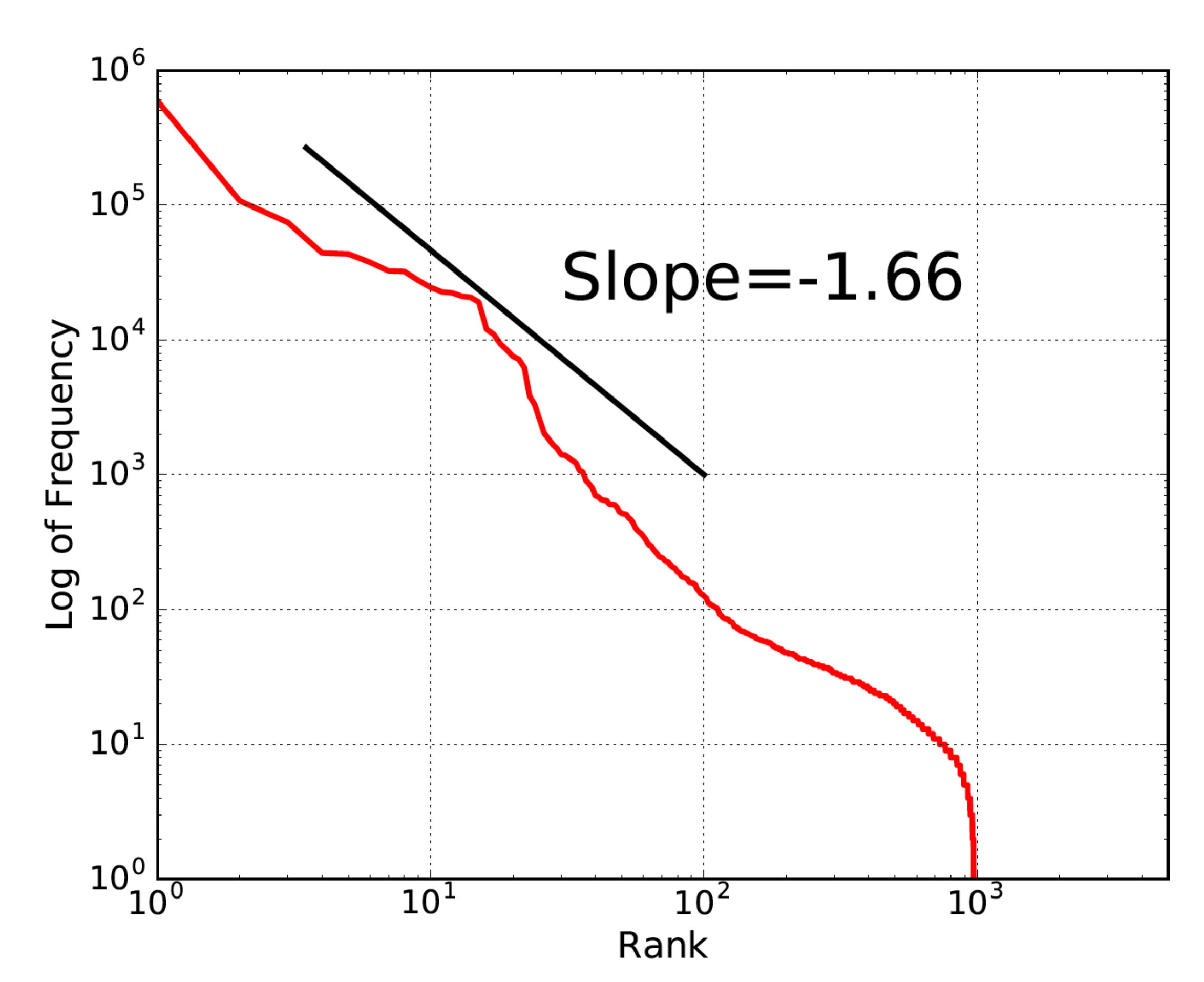}
     \caption[``Smart'' premium fraudster follow counts]{P1 (``smart'') follow count}
     \label{fig:rf_p_smart_followers}
    \end{subfigure}
    &
    \begin{subfigure}[t]{0.43\textwidth}
     \centering
     \includegraphics[width=\textwidth]{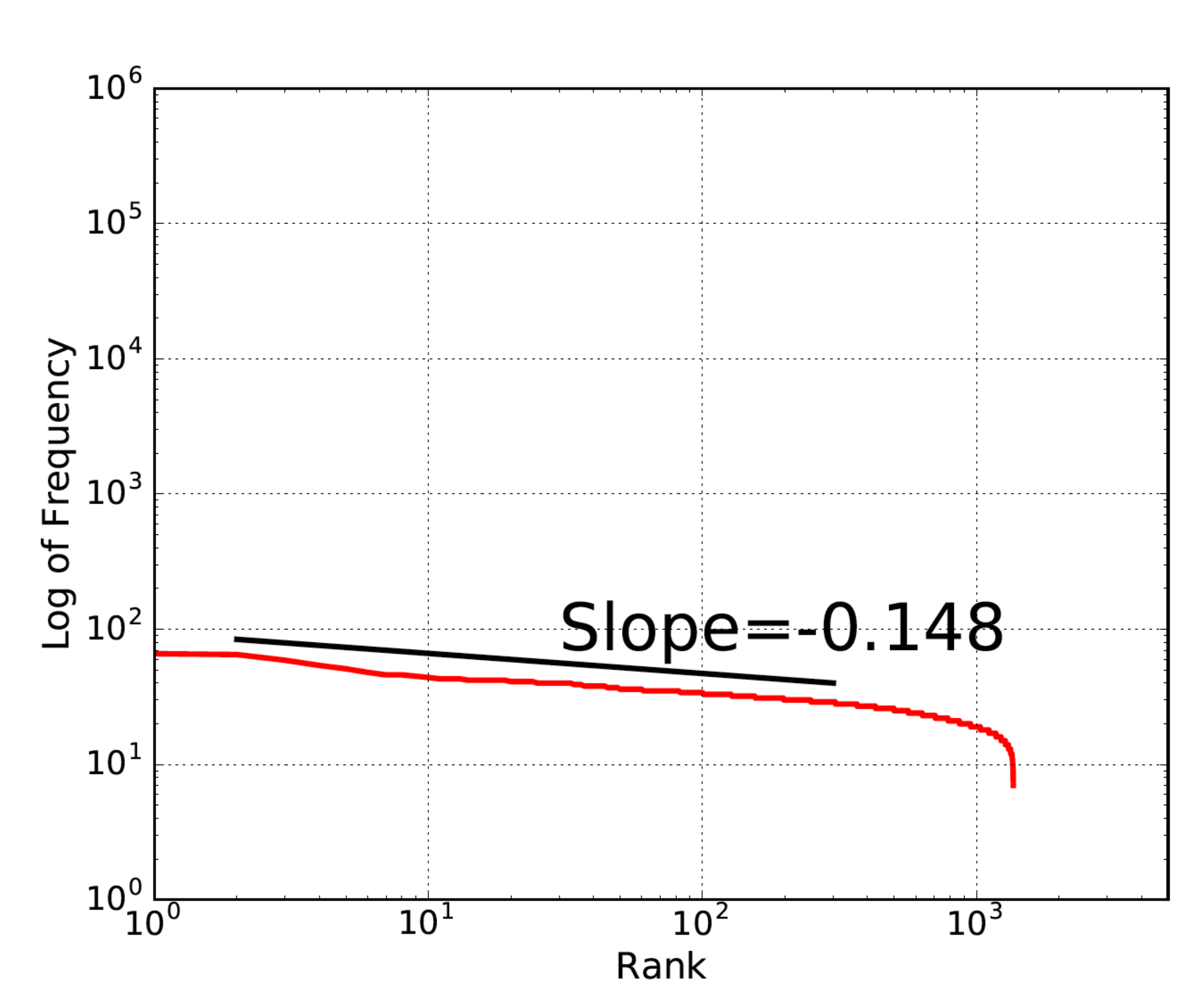}
     \caption[``Na\"{i}ve'' premium fraudster follow counts]{P2 (``na\"{i}ve'') follow count}
     \label{fig:rf_p_naive_followers}
    \end{subfigure}
   \end{tabular}
  \caption[Differences across user types' follower action counts]{\textbf{Rank-frequency plots reveal different patterns in follower counts of various follower types.}  Note that genuine follower counts in (a) reflect traditional power-law behavior with a common exponent ($\sim1.2$) and are linear in log-log scale.  Freemium counts in (b) fit similarly, despite with a slightly lower exponent ($\sim1.15$).  Comparatively, ``smart'' premium counts in (c) fit a power law but with much higher exponents ($\sim1.66$). Interestingly, we find that ``na\"{i}ve'' premium followers do fit a power law, but have unnaturally low exponents ($\sim.148$) due to their low entropy and highly concentrated, robotic behavior.}
  \label{fig:rfs}
\end{figure*}

\devumi and \twitterboost have much lower entropy for action counts (favorites, followers, friends, lists and statuses) compared to freemium providers.  \fastfollowerz also exhibits lower entropy.  As Figure \ref{fig:crown_attributes} shows, there is even more variation between premium providers.  Figure \ref{fig:crown_attributes} shows that \intertwitter (P1 ``smart'') follower counts are disparate and closer to genuine users' entropy, unlike other premium fraudsters (P2 ``na\"{i}ve'') who behave robotically.  Comparatively, freemium followers have lower follower count entropy compared to genuine ones, which is intuitive as while the freemium follows are real accounts, their follower counts are not independent from each other due to the follows traded between themselves.  Figure \ref{fig:rfs} shows the rank-frequency plots for follower counts for various follower types.  The plots substantiate our observations on entropy, and also show that different user types exhibit differences with regards to power-law fit, which is expected for skewed distributions on social networks. While entropy values in this paper are computed empirically using the samples from Table \ref{tbl:egonet_stats}, accounts on real networks have varying follower counts, leading to different entropy estimates even when drawn from the same distribution.  Fortunately, we can intimately relate sample size and entropy of a power-law distribution in a closed form using the Euler-Maclaurin formula as below:
\begin{lemma}[Power-Law Entropy]
 The entropy $H$ of a size $|V|$ sample from a PL distribution $P(r) = C \cdot r^{-a}$ is given by:
 \begin{displaymath}
  H \approx -\frac{C \cdot \log_2(C) \cdot (|V|^{1-\alpha} - 1)}{1 - \alpha} + \frac{\alpha \cdot C \cdot \left(-|V|^{1 - \alpha} \cdot \left(\left(\alpha - 1\right) \cdot \ln(|V|) + 1\right) + 1\right)}{(\alpha - 1)^2 \cdot \ln(2)} 
 \end{displaymath}
  
  \begin{proof} 
{
\openup-1\jot
 \begin{align*}
 H &= -\sum\limits_{r=1}^{|V|} Cr^{-\alpha} \cdot \log_2 C \cdot r^{-\alpha} \\
    &\approx -\int_{1}^{V} C \cdot r^{-\alpha}  \cdot \log_2 C dr + \alpha  \cdot C \int_{1}^{V} r^{-\alpha}  \cdot  \log_2 r dr \\
    &\approx -\frac{C  \cdot \log_2 C  \cdot r^{1 - \alpha}}{1 - \alpha} \Big|_{1}^{V} + \alpha  \cdot C \left( \frac{-r^{1-\alpha}\left(\left(\alpha - 1\right)  \cdot  \ln(r) + 1\right)}{(\alpha - 1)^2 \cdot \ln(2)} \Big|_{1}^{V}  \right) \\
    &\approx -\frac{C \cdot \log_2(C) \cdot (|V|^{1-\alpha} - 1)}{1 - \alpha} + \\
    &\quad \quad  \frac{\alpha \cdot C \cdot \left(-|V|^{1 - \alpha} \cdot \left(\left(\alpha - 1\right) \cdot \ln(|V|) + 1\right) + 1\right)}{(\alpha - 1)^2} 
    \tag*{\qedhere}
 \end{align*}}
\end{proof}
\noindent where $C = {1}/{H_{|V|,\alpha}}$ (inverse of the $V^{th}$ harmonic number of order $\alpha$).
\end{lemma}

This estimate enables us to gauge how close varying-sized samples are to the original power-law.  This is especially useful for practitioners aiming to gauge what the entropy of an account's followers' attributes theoretically \emph{should be} according to the number of followers assuming a given power-law fit, versus the empirical estimate.  If these are not close, one can deduce that the account's followers \emph{do not} obey the expected power-law fit and therefore may be suspicious.  This procedure is computationally more efficient and likely more accurate than fitting a separate power-law for each of the attributes across followers of each account.

We noticed similar patterns in entropy for status and favorite counts as well.  The lower entropy of action counts characteristic of premium providers stems from the variety of options premium providers have for Twitter engagement -- in addition to fake followers, the premium providers also offer fake retweets and favorites services.  Thus, premium providers are incentivized to reuse accounts for multiple types of fraud, and when done na\"{i}vely result in high synchrony in ``serviceable'' attributes.

\begin{figure}[t]
 \centering
 \begin{subfigure}[t]{0.48\textwidth}
 \includegraphics[width=\textwidth]{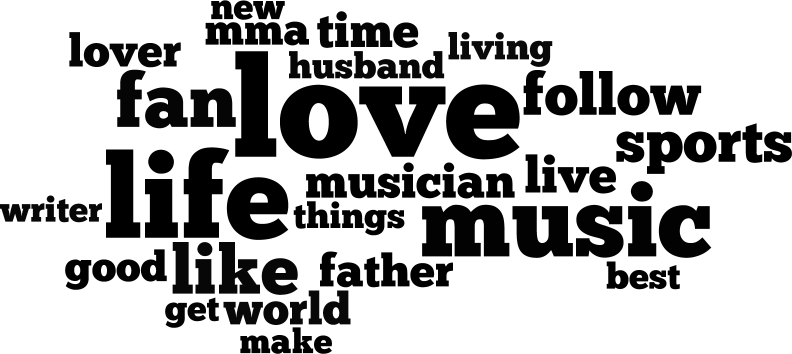}
 \caption[Premium fraudster word use]{Premium}
 \label{fig:premium_wordle}
 \end{subfigure} 
 \hfill
 \begin{subfigure}[t]{0.48\textwidth}
 \includegraphics[width=\textwidth]{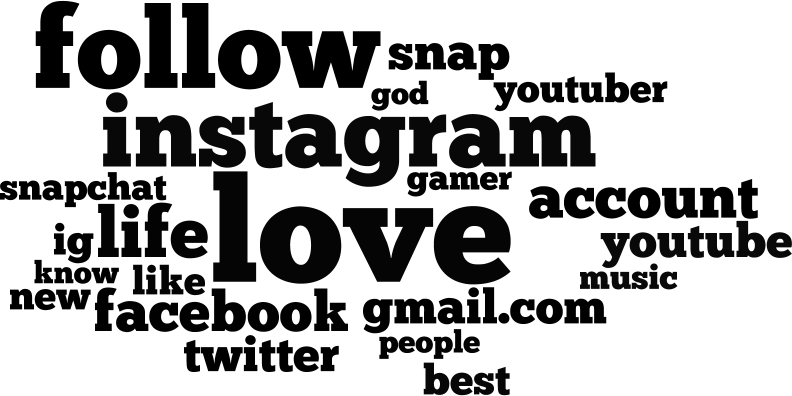}
 \caption[Freemium fraudster word use]{Freemium}
 \label{fig:freemium_wordle}
 \end{subfigure}
 \caption[Comparison of descriptive word-use between fraud types]{\textbf{Freemium followers have social media (Facebook, Instagram, Snapchat) focused descriptions (right), whereas premium followers have wordy descriptions (left).}}
 \label{fig:wordles}
\end{figure}

\vspace{1mm}
\subsubsection{User Settings}
\fastfollowerz, \devumi and \twitterboost all have near 0 geolocation, language, and tweet protection entropy. Of these, all \devumi and \twitterboost accounts use the US English language setting, have geolocation disabled and do not protect tweets. \fastfollowerz has a slightly higher language entropy of .06, but we found that all \fastfollowerz accounts were either using US or GB English, suggesting a heavy premium bias for English accounts. We also found that premium followers almost entirely have USA timezones.  ``Smart'' \intertwitter followers' high language entropy from Figure \ref{fig:crown_attributes} suggests an aim to better camouflage user attributes compard to the ``na\"{i}ve'' providers.   Given that \intertwitter also has some verified accounts, we hypothesize that the accounts may be hijacked ones.  This is in contrast with freemium providers, which have much higher frequency of enabled geolocation, variance in language and protected tweets.  Figure \ref{fig:crown_attributes} also shows that freemium followers tend to appear similar to genuine ones as they are otherwise real user accounts.  However, we find that freemium followers have higher language entropy than genuine ones, as freemium followers are spread over many languages whereas genuine followers tend to disproportionately speak their followee's language (i.e. if a user speaks Spanish, most of his followers speak Spanish).  
\vspace{1mm}

Furthermore, all 4 freemium providers and \twitterboost/\devumi have extremely similar attribute entropy over their fake followers respectively, further substantiating Insight \ref{ins:collusion}.

In addition to the attributes reported in Table \ref{tbl:attrib_entropy}, we also studied the 160-character user description field.  The description field essentially contains the high-level summary of what the user aims to appear as to other Twitter users, and is thus interesting to analyze.  
We ask: what, if any, are the differences between freemium and premium follower descriptions? 

Figure \ref{fig:wordles} shows two wordclouds, aggregated over description text across all premium and freemium followers respectively.  Font size corresponds to relative frequency in the text.  For clarity, we remove common stopwords. We arrive at the following insight:

\begin{insight}[Clout vs. About]
\label{ins:cloutvsabout}
Freemium followers tend to have descriptions focusing on social media clout, 
whereas premium followers tend to talk about themselves.
\end{insight}

Figure \ref{fig:premium_wordle} (premium), has words like ``musician,'' ``lover,'' ``writer'' and ``sports'', corresponding to descriptive personal details -- these are likely copied from genuine users.
Conversely, Figure \ref{fig:freemium_wordle} (freemium) has terms like ``snapchat,'' ``youtube,'' and ``instagram'', as these users try to increase clout by advertising their other, real social media pages, i.e., ``\emph{follow me on snapchat}.''

\begin{figure}[t!]
	\centering
	\includegraphics[width=0.7\textwidth]{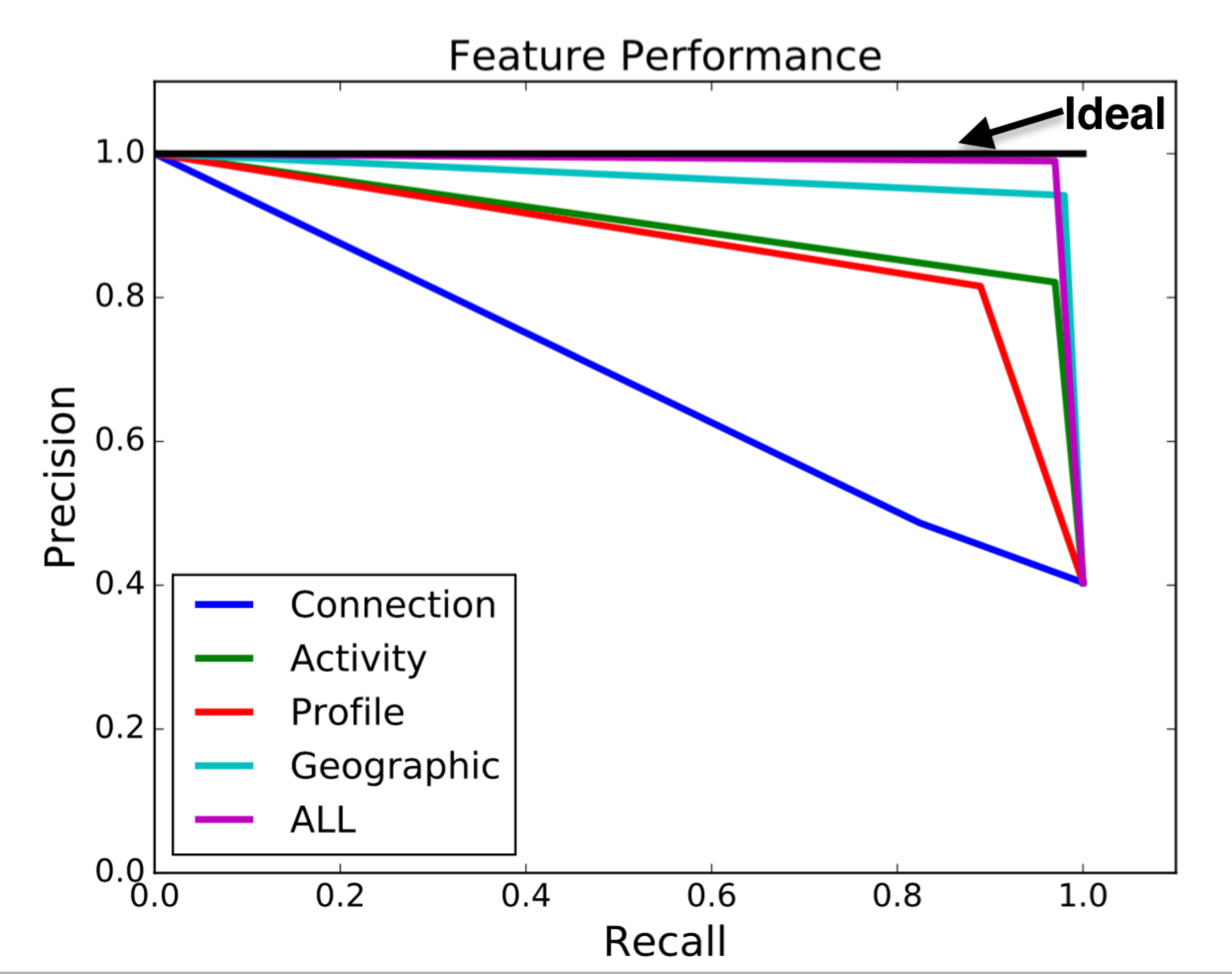}
	\caption[Entropy-based features show strong classification performance]{\textbf{Leveraging all features together gives the best detection performance.}}
	\label{figure:prec_recall}
\end{figure}

\section{Assessing Discriminative Power of Entropy Features}
\label{sec:disc_power}
Thus far, we have highlighted a number of distributional differences between fraudulent and genuine users. Can we leverage these differences to discriminate user behaviors? In this section, we evaluate a number of attribute features on their discriminative power in a supervised setting.

We classified the engineered entropy features from Table~\ref{tbl:attrib_entropy} into the following groups based on feature type:

\vspace{1mm}
\begin{compactitem}
\item \textit{Connection}: \# Followers, \# Friends
\item \textit{Activity}: \# Statuses, \# Lists, \# Favorites
\item \textit{Profile}: Default Profile (and Image), Verified, Created
\item \textit{Geography}: Language, UTC
\item \textit{All}: the union of all above features
\end{compactitem} 

\vspace{1mm}
Note that while we nominally refer to these features as above, they refer to the \emph{entropy of the feature over account followers}, rather than raw values of the account itself.

We evaluate these features using binary classification (genuine vs. fraudulent) as is traditionally done in practice. We use a Support Vector Machine (SVM) with radial basis function (RBF) kernel and $10$-fold cross validation as the classifier of choice, but any out-of-box classification method could be used. Our carefully assembled ground-truth dataset consists of 307 fraudulent users and 200 genuine users, whose features are computed over their followers. The fraudulent accounts are a combination of premium and freemium honeypots as well as accounts whose profiles have been listed on freemium providers' websites as users of the service.  We define our fraudulent set over this multitude of account types with various properties in order to demonstrate generality. The genuine accounts belong to well-known academics in machine learning and data mining.  We avoid using randomly sampled Twitter users, as previous works have shown a non-trivial amount of fake accounts on Twitter which may excessively corrupt our ground-truth genuine set. In practice, getting additional ground-truth labels is a very costly endeavor and requires careful manual inspection for each individual case.
      
Figure \ref{figure:prec_recall} shows the relative performance of our feature groups in terms of overall precision and recall. We notice that \emph{Connection} features perform comparatively poorly, \emph{Profile} and \emph{Activity} features perform better, \emph{Geography} performs even better, and the combination \emph{All} performs near-ideal with .98 precision and .95 recall (much higher recall than supervised approaches which use raw account features for Twitter spam classification \cite{mccord2011spam}). Thus, we conclude that our proposed entropy features are highly reliable in discerning genuine from fraudulent users.  The added benefit of using the entropy-based features is that it is much harder to control for from the fraudster's perspective -- this is because while the fraudster has significant control over his own account's properties, he has limited ability to influence who follows him.  

\section{Discussion}



The analysis in this work has a number of important implications on fraud detection in practice.  We detail these below.

\vspace{1mm}
\noindent \textbf{Multimodal Detection:} Using individual signatures to find one type of fraud tends to be at the expense of finding other types. For example, clique detection primarily focuses on freemium fraud, whereas bipartite core detection focuses on premium fraud. Using complementary methods is a promising strategy.

\vspace{1mm}
\noindent \textbf{Importance of Time:} Varying account reuse policies makes temporal granularity an important consideration in graph-based fraud detection. While analysis on a low granularity graph can reveal dense fraudulent structure in frequent reuse regimes, it may never do so for low reuse regimes.  Higher granularity can be useful in these cases. 

\vspace{1mm}
\noindent \textbf{Deceptive Account Attributes:} Using individual account attributes to label fraudsters is of limited use. Our work suggests that most freemium fraudsters are actually real users with real profile attributes -- they may be resistant to such detection schemes. Conversely, leveraging an account's follower's attributes shows promise in bridging this gap.

\vspace{1mm}
\noindent \textbf{Total vs. Partial Fraud:} Different types of fraud may call for different penalties.  While the implication ``has one fake link $\rightarrow$ has all fake links'' seems true for premium fraudsters, it is not for freemium ones.  Removing fake links vs. suspending fake accounts is a promising way to penalize such fraudsters and minimize false positives.

\vspace{1mm}
The need for multimodal anti-fraud mechanisms suggests a shift in the detection paradigm from drawing a two-class boundary between genuine and ``one-hat-fits-all'' fraudulent users, to a more complex multiclass boundary between genuine, premium fraudulent, freemium fraudulent, and other fraud types which may be discovered in the future.
\section{Conclusion}

 In this work, we aimed to study the nature of modern link fraud regimes.  To this end, we setup honeypot accounts on Twitter, purchased fake followers for them from a variety of fraud-provoding services, and carefully instrumented a data scraping process to capture their behaviors. Specifically, we studied the local network connectivity of fake followers via the egonet and proposed boomerang networks, as well as attribute distributions over profile features and account actions.  Our analyses showed that there are multiple types of link fraud (we discover at least two: \emph{freemium} and \emph{premium}) with varying behaviors regarding internal and external network connectivity, disparity in attribute homogeneity across followers, and differences in descriptive word-usage in Twitter bios. Furthermore, we found fascinating evidence that service providers have varying types of account-reuse policies and seem to collude with each other on a number of fronts.  Furthermore, we proposed the use of first-order entropy features taken across account followers' attributes to discern fraudulent from genuine accounts, and showed that these features were able to attain near-perfect F1 score on our ground-truth dataset.  Holistically, our work offers several implications for practical fraud detection including multimodality of fraud behaviors, the importance of temporally sensitive algorithms, usefulness of first-order versus zeroth-order features, and disadvantages of account-based versus link-based fraud targeting.

\bibliographystyle{abbrv}
\bibliography{bib/neil,bib/hemank} 

\begin{thebibliography}{10}

\bibitem{Aggarwal:2015}
A.~Aggarwal and P.~Kumarguru.
\newblock What they do in shadows: Twitter underground follower market.
\newblock In {\em PST}, 2015.

\bibitem{Benevenuto:2010}
F.~Benevenuto, G.~Magno, T.~Rodrigues, and V.~Almeida.
\newblock Detecting spammers on twitter.
\newblock In {\em CEAS}, 2010.

\bibitem{beutel2013copycatch}
A.~Beutel, W.~Xu, V.~Guruswami, C.~Palow, and C.~Faloutsos.
\newblock Copycatch: stopping group attacks by spotting lockstep behavior in
  social networks.
\newblock In {\em WWW}, pages 119--130. International World Wide Web
  Conferences Steering Committee, 2013.

\bibitem{Cai:2012}
Z.~Cai and C.~Jermaine.
\newblock The latent community model for detecting {Sybils} in social networks.
\newblock In {\em NDSS}, Feb. 2012.

\bibitem{cao2014uncovering}
Q.~Cao, X.~Yang, J.~Yu, and C.~Palow.
\newblock Uncovering large groups of active malicious accounts in online social
  networks.
\newblock In {\em CCS}, pages 477--488. ACM, 2014.

\bibitem{freeman2013using}
D.~M. Freeman.
\newblock Using naive bayes to detect spammy names in social networks.
\newblock In {\em {AISec}}, pages 3--12. ACM, 2013.

\bibitem{Gao:2010}
H.~Gao, J.~Hu, C.~Wilson, Z.~Li, Y.~Chen, and B.~Y. Zhao.
\newblock Detecting and characterizing social spam campaigns.
\newblock In {\em SIGCOMM}, IMC '10, pages 35--47, New York, NY, USA, 2010.
  ACM.

\bibitem{grier2010spam}
C.~Grier, K.~Thomas, V.~Paxson, and M.~Zhang.
\newblock @ spam: the underground on 140 characters or less.
\newblock In {\em CCS}, pages 27--37. ACM, 2010.

\bibitem{Gupta:13}
A.~Gupta, H.~Lamba, and P.~Kumaraguru.
\newblock \$1.00 per rt \#bostonmarathon \#prayforboston: Analyzing fake
  content on twitter.
\newblock In {\em Proceedings of the eCrime Researchers Summit (eCRS)}, pages
  1--12. IEEE, 2013.

\bibitem{Gupta:2013}
A.~Gupta, H.~Lamba, P.~Kumaraguru, and A.~Joshi.
\newblock Faking sandy: Characterizing and identifying fake images on twitter
  during hurricane sandy.
\newblock In {\em WWW}, WWW '13 Companion, pages 729--736, New York, NY, USA,
  2013. ACM.

\bibitem{jiang2014catchsync}
M.~Jiang, P.~Cui, A.~Beutel, C.~Faloutsos, and S.~Yang.
\newblock Catchsync: catching synchronized behavior in large directed graphs.
\newblock In {\em KDD}, pages 941--950. ACM, 2014.

\bibitem{Lee:2010}
K.~Lee, J.~Caverlee, and S.~Webb.
\newblock Uncovering social spammers: Social honeypots + machine learning.
\newblock In {\em SIGIR}, SIGIR '10, pages 435--442, New York, NY, USA, 2010.
  ACM.

\bibitem{mccord2011spam}
M.~Mccord and M.~Chuah.
\newblock Spam detection on twitter using traditional classifiers.
\newblock In {\em ICATC}, pages 175--186. Springer, 2011.

\bibitem{Motoyama:2010}
M.~Motoyama, K.~Levchenko, C.~Kanich, D.~McCoy, G.~M. Voelker, and S.~Savage.
\newblock Re: Captchas: Understanding captcha-solving services in an economic
  context.
\newblock In {\em USENIX Security}, USENIX Security'10, pages 28--28, Berkeley,
  CA, USA, 2010. USENIX Association.

\bibitem{TwitterWebImpact:13}
N.~Perloth.
\newblock Fake twitter followers become multimillion-dollar business, April
  2013.
\newblock [Online; posted 5-April-2013].

\bibitem{prakash2010eigenspokes}
B.~A. Prakash, A.~Sridharan, M.~Seshadri, S.~Machiraju, and C.~Faloutsos.
\newblock Eigenspokes: Surprising patterns and scalable community chipping in
  large graphs.
\newblock In {\em PAKDD}, pages 435--448. Springer, 2010.

\bibitem{shah2014spotting}
N.~Shah, A.~Beutel, B.~Gallagher, and C.~Faloutsos.
\newblock Spotting suspicious link behavior with fbox: An adversarial
  perspective.
\newblock In {\em ICDM}, pages 959--964. IEEE, 2014.

\bibitem{Stringhini:2010}
G.~Stringhini, C.~Kruegel, and G.~Vigna.
\newblock Detecting spammers on social networks.
\newblock In {\em ACSAC}, ACSAC '10, pages 1--9, New York, NY, USA, 2010. ACM.

\bibitem{Stringhini:2013}
G.~Stringhini, G.~Wang, M.~Egele, C.~Kruegel, G.~Vigna, H.~Zheng, and Y.~B.
  Zhao.
\newblock Follow the green: Growth and dynamics in twitter follower markets.
\newblock In {\em SIGMETRICS}, 2013.

\bibitem{Thomas:2014}
K.~Thomas, D.~Iatskiv, E.~Bursztein, T.~Pietraszek, C.~Grier, and D.~McCoy.
\newblock Dialing back abuse on phone verified accounts.
\newblock In {\em CCS}, CCS '14, pages 465--476, New York, NY, USA, 2014. ACM.

\bibitem{Thomas:2013}
K.~Thomas, D.~McCoy, C.~Grier, A.~Kolcz, and V.~Paxson.
\newblock Trafficking fraudulent accounts: The role of the underground market
  in twitter spam and abuse.
\newblock In {\em USENIX Security}, SEC'13, pages 195--210, Berkeley, CA, USA,
  2013. USENIX Association.

\bibitem{Wang:2012}
G.~Wang, C.~Wilson, X.~Zhao, Y.~Zhu, M.~Mohanlal, H.~Zheng, and B.~Y. Zhao.
\newblock Serf and turf: Crowdturfing for fun and profit.
\newblock In {\em WWW}, WWW '12, pages 679--688, New York, NY, USA, 2012. ACM.

\bibitem{yu2006sybilguard}
H.~Yu, M.~Kaminsky, P.~B. Gibbons, and A.~Flaxman.
\newblock Sybilguard: defending against sybil attacks via social networks.
\newblock {\em SIGCOMM}, 36(4):267--278, 2006.

\end{thebibliography}
\end{document}